\documentclass[a4paper,11pt]{article}
\pdfoutput=1 
\usepackage{jcappub}
\usepackage{graphicx}
\pdfoutput=1 
\usepackage{color}
\usepackage{makecell}
\usepackage{afterpage}
\usepackage[dvipsnames]{xcolor}
\usepackage{multirow}
\usepackage{cleveref}
\usepackage[normalem]{ulem}		
\usepackage{xspace}
\usepackage{subcaption}
\usepackage{booktabs}
\usepackage{csquotes}

\usepackage{centernot}
\usepackage{mathtools}
\usepackage{stmaryrd}
\usepackage{float}
\usepackage{listings}

\newcommand\dm{{\mathrm{DM}}}

\newcommand\dark{{\mathrm{dark}}}

\title{A step in the right direction? Analyzing the Wess Zumino Dark Radiation solution to the Hubble tension}

\author[1]{Nils Schöneberg}
\author[2]{and Guillermo Franco Abellán}

\emailAdd{nils.science@gmail.com}
\affiliation[1]{Dept. F\'isica Qu\`antica i Astrof\'isica, Institut de Ci\`encies del Cosmos (ICCUB), Facultat de F\'isica, Universitat de Barcelona (IEEC-UB), Mart\'i i Franqu\'es, 1, E08028 Barcelona, Spain\\}

\affiliation[2]{Laboratoire Univers \& Particules de Montpellier (LUPM), CNRS \& Universit\'e de Montpellier (UMR-5299),Place Eug\`ene Bataillon, F-34095 Montpellier Cedex 05, France}
\emailAdd{guillermo.franco-abellan@umontpellier.fr}

\abstract{The Wess Zumino Dark Radiation (WZDR) model first proposed in \cite{Aloni:2021eaq} shows great promise as a well-motivated simple explanation of the Hubble tension between local and CMB-based measurements. In this work we investigate the assumptions made in the original proposal and confront the model with additional independent data sets. We show that the original assumptions can have an impact on the overall results but are usually well motivated. We further demonstrate that the preference for negative $\Omega_k$ observed in Planck data remains at a similar level as for the $\Lambda$CDM model, while the $A_L$ tension is slightly increased. Furthermore, the tension between Planck data for $\ell < 800$ and $\ell \geq 800$ is significantly reduced for the WZDR model.
The Planck-independent data sets show slightly more permissive bounds on the Hubble parameter, allowing the tension to be further reduced to $2.1 \sigma$ (CMB-independent) or $1.9\sigma$ (ACT+WMAP). However, no combination shows a large preference for the presence of WZDR. We also investigate whether additional dark radiation -- dark matter interactions can help in easing the $S_8$ tension as well. Assuming all of the dark matter to be interacting and a temperature-independent scattering rate, we find that the CMB data are too restrictive on this additional component as to allow a significant decrease in the clustering.}

\begin{document}

\def\aj{\ref@jnl{AJ}}                   
\def\actaa{\ref@jnl{Acta Astron.}}      
\def\araa{\ref@jnl{ARA\&A}}             
\def\apj{\ref@jnl{ApJ}}                 
\def\apjl{\ref@jnl{ApJ}}                
\def\apjs{\ref@jnl{ApJS}}               
\def\ao{\ref@jnl{Appl.~Opt.}}           
\def\apss{\ref@jnl{Ap\&SS}}             
\def\aap{\ref@jnl{A\&A}}                
\def\aapr{\ref@jnl{A\&A~Rev.}}          
\def\aaps{\ref@jnl{A\&AS}}              
\def\azh{\ref@jnl{AZh}}                 
\def\baas{\ref@jnl{BAAS}}               
\def\bac{\ref@jnl{Bull. astr. Inst. Czechosl.}}
\def\caa{\ref@jnl{Chinese Astron. Astrophys.}}
\def\cjaa{\ref@jnl{Chinese J. Astron. Astrophys.}}
\def\icarus{\ref@jnl{Icarus}}           
\def\jcap{\ref@jnl{J. Cosmology Astropart. Phys.}}
\def\jrasc{\ref@jnl{JRASC}}             
\def\memras{\ref@jnl{MmRAS}}            
\def\mnras{\ref@jnl{MNRAS}}             
\def\na{\ref@jnl{New A}}                
\def\nar{\ref@jnl{New A Rev.}}          
\def\pra{\ref@jnl{Phys.~Rev.~A}}        
\def\prb{\ref@jnl{Phys.~Rev.~B}}        
\def\prc{\ref@jnl{Phys.~Rev.~C}}        
\def\prd{\ref@jnl{Phys.~Rev.~D}}        
\def\pre{\ref@jnl{Phys.~Rev.~E}}        
\def\prl{\ref@jnl{Phys.~Rev.~Lett.}}    
\def\pasa{\ref@jnl{PASA}}               
\def\pasp{\ref@jnl{PASP}}               
\def\pasj{\ref@jnl{PASJ}}               
\def\rmxaa{\ref@jnl{Rev. Mexicana Astron. Astrofis.}}%
\def\qjras{\ref@jnl{QJRAS}}             
\def\skytel{\ref@jnl{S\&T}}             
\def\solphys{\ref@jnl{Sol.~Phys.}}      
\def\sovast{\ref@jnl{Soviet~Ast.}}      
\def\ssr{\ref@jnl{Space~Sci.~Rev.}}     
\def\zap{\ref@jnl{ZAp}}                 
\def\nat{\ref@jnl{Nature}}              
\def\iaucirc{\ref@jnl{IAU~Circ.}}       
\def\aplett{\ref@jnl{Astrophys.~Lett.}} 
\def\apspr{\ref@jnl{Astrophys.~Space~Phys.~Res.}}
\def\bain{\ref@jnl{Bull.~Astron.~Inst.~Netherlands}} 
\def\fcp{\ref@jnl{Fund.~Cosmic~Phys.}}  
\def\gca{\ref@jnl{Geochim.~Cosmochim.~Acta}}   
\def\grl{\ref@jnl{Geophys.~Res.~Lett.}} 
\def\jcp{\ref@jnl{J.~Chem.~Phys.}}      
\def\jgr{\ref@jnl{J.~Geophys.~Res.}}    
\def\jqsrt{\ref@jnl{J.~Quant.~Spec.~Radiat.~Transf.}}
\def\memsai{\ref@jnl{Mem.~Soc.~Astron.~Italiana}}
\def\nphysa{\ref@jnl{Nucl.~Phys.~A}}   
\def\physrep{\ref@jnl{Phys.~Rep.}}   
\def\physscr{\ref@jnl{Phys.~Scr}}   
\def\planss{\ref@jnl{Planet.~Space~Sci.}}   
\def\procspie{\ref@jnl{Proc.~SPIE}}   

\let\astap=\aap
\let\apjlett=\apjl
\let\apjsupp=\apjs
\let\applopt=\ao

\newcommand{\nilsc}[1]{\textbf{\color{ForestGreen}[NS:~#1]}}
\newcommand{\nils}[1]{{\color{ForestGreen}#1}}
\newcommand{\nilsr}[2]{{\color{ForestGreen}\sout{#1}#2}}
\newcommand{\der}[2]{\frac{\mathrm{d}#1}{\mathrm{d}#2}}

\newcommand{\guillec}[1]{\textbf{\color{Plum}[GFA:~#1]}}
\newcommand{\guille}[1]{{\color{Plum}#1}}
\newcommand{\guiller}[2]{{\color{Plum}\sout{#1}#2}}

\maketitle

\section{Introduction} \label{sec:intro}

 The Hubble tension refers to the inconsistency between the locally measured Hubble parameter and its value as predicted from observations of the Cosmic Microwave Background (CMB), assuming $\Lambda$CDM. Despite the multitude of local measurements based on different anchors, standard candles, and analysis methodologies \cite{Riess:2021jrx,Brout:2022vxf,Soltis:2020gpl,Anand:2021sum,Philcox:2020vvt}, as well as the multitude of CMB based measurements from Planck \cite{Planck:2018vyg}, ACT \cite{ACT:2020gnv}, and SPT \cite{SPT-3G:2021eoc}, the tension between local and CMB-based inferences of the Hubble parameter remains at a level ranging around 4-$5 \sigma$. See also \cite{Schoneberg:2021qvd,DiValentino:2021izs,Verde:2019ivm} for some recent reviews. See also \cite{Colgain:2022rxy,Colgain:2022nlb,Romano:2016utn,Dainotti:2021pqg} for a different perspective on the Hubble tension.

Under the assumption that there are no major systematic errors in the corresponding data sets, it is natural to question the model used to infer the Hubble parameter from the CMB. However, the inference of the Hubble parameter turns out to be surprisingly robust to modifications of the underlying cosmological model, and many proposed solutions had to resort to some degree of fine-tuning or introduce a multitude of new ingredients \cite{Schoneberg:2021qvd}. As such, finding a simple model that eases the Hubble tension has become one of the most difficult challenges to cosmological model builders.

Recently, in \cite{Aloni:2021eaq} the authors have demonstrated  that a model of Wess-Zumino dark radiation appears to be capable of at least partially resolving the Hubble tension, reducing it to a significance of around $2.7\sigma$ only. Built on a relatively straightforward and well-motivated extension of the standard $\Lambda$CDM model, this model appears to surpass many of the challenges that previous models based on dark radiation have faced.

In particular, the increased Silk damping and other perturbative effects (like the \enquote{drag effect}) \cite{Bashinsky:2003tk,Lesgourgues:2013sjj,Baumann:2015rya} usually spoil the excellent agreement of the high-$\ell$ polarization measurements of the CMB when increasing the amount of dark radiation present in the universe \cite{Bernal:2016gxb,Planck:2018vyg,DiValentino:2019dzu,Schoneberg:2021qvd}. If the dark radiation is not assumed to be free-streaming, but instead strongly self interacting, it has a much smaller impact on the CMB observables and thus allows for larger $\Delta N_\mathrm{eff}$ and $H_0$ to be inferred. This is due to the additional clustering of the radiation, which  partially mitigates the increased Silk damping. However, this model is still tightly constrained by Planck polarization measurements \cite{Aloni:2021eaq,Schoneberg:2021qvd}. 

If the dark radiation were to only influence lower $\ell$ modes, then such a model could still cause a shift in the sound horizon while avoiding the penalty from the high-$\ell$ polarization. This surplus of additional dark radiation at a given time can be readily generated by the decay of a more massive state\footnote{Indeed, this is the same mechanism responsible for recombination, big bang nucleosynthesis, and most likely also baryogenesis.}. This is precisely the setup of the WZDR model, where two species (one massive and one massless) at early times coexist with only a small dark radiation contribution. As the massive particle becomes non-relativistic, it transfers its energy into heating the remaining massless species, overall increasing the abundance of dark radiation. By itself, the added dark radiation already decreases the sound horizon $r_s$ and through its well-known degeneracy with $H_0$ for constant sound horizon angle $\theta_s$ also increases the Hubble parameter. Additionally, the step-like nature of the impact of the WZDR on CMB scales both allows for larger amounts of radiation through the avoidance of the strict polarization bounds, as well as a minor ($\sim 1\sigma$, see \cite{Aloni:2021eaq}) upward shift in $\theta_s$\,, which additionally contributes to the shift in $H_0$\,.

Given the interest in such a simply motivated model with a strong impact on the Hubble tension, it is natural to further investigate the underlying assumptions and impacts on the cosmological observables. In this work, we check which choices made during the analysis are beneficial or detrimental to the ability of the model to ease the Hubble tension. Further, we show how other well known tensions of the $\Lambda$CDM cosmological model \cite{Perivolaropoulos:2021jda,Planck:2018vyg} (such as the lensing tension of the $A_L$ parameter, the curvature preference, and the difference in parameter inference for $\ell < 800$ and $\ell \geq 800$) are impacted by this model. We further investigate whether the ability to ease the Hubble tension remains when confronting this model with different alternative datasets. Finally, we check if an extension with interacting dark matter can simultaneously ease the $S_8$ tension with weak lensing observables \cite{DES:2021wwk,KiDS:2020suj,Planck:2018vyg}. While the $S_8$ tension is of smaller significance and could be statistical in nature \cite{Nunes:2021ipq}, its persistence among several low-redshift probes \cite{Abdalla:2022yfr} remains very intriguing, and could be pointing to new physics beyond the $\Lambda$CDM paradigm.

In \cref{sec:model} we introduce our updated and extended description of the WZDR model originally proposed by \cite{Aloni:2021eaq}. In \cref{sec:results} we answer the questions posed above, and we summarize our findings in \cref{sec:concl}.
\section{Description of the model} \label{sec:model}

Despite its complex impact on the CMB, the WZDR model can be motivated using only a simple supersymmetric model involving a single superfield in a cubic superpotential, as discussed in \cite{Aloni:2021eaq}. The corresponding Lagrangian involves two massless dark species; a strongly self-interacting boson, and a strongly interacting fermion. A mass term for the boson could be subsequently generated from supersymmetry breaking or from gravitational effects. However, the actual dynamics of this model can be described quite independently of this motivation. As such, other theoretical foundations beyond supersymmetry are also worth investigating. In view of a general description, we assume the existence of a massless species $\phi$ and a massive species $\xi$ with mass $m_\xi$ which have strong interactions with each other.

In general, the cosmological evolution of the WZDR model proceeds along three major phases. Initially, both $\phi$ and $\xi$ are in thermal and chemical equilibrium. At this point both the decay of the massive species $\xi$ into the massless species $\phi$ as well as the inverse decay are efficient. As the universe expands and the temperature of each species is diluted, the average momentum of the massless species $\phi$ will soon reach the mass-threshold $m_\xi$ of the massive species. At this point, the inverse decay process quickly becomes inefficient. This in itself would not have a big impact on the cosmological evolution. However, simultaneously the massive species becomes non-relativistic at this point. Thus its number and energy densities become exponentially suppressed compared to the massless species. In the end, the massless species continues evolving with a slightly increased number density and temperature due to the decay, while the massive species has become completely negligible\footnote{This process of non-relativistic transition is of course similar to that of electron positron annihilation, except for the lack of imbalance of positive and negative charges. Another similar situation arises in models of active neutrinos interacting with the majoron (a pseudo-Goldstone boson arising from the spontaneous breaking of global lepton number), which were proposed as a resolution to the $H_0$ tension in \cite{Escudero:2019gvw, Escudero:2021rfi}. In these models, inverse neutrino decays thermalize with the majoron population (which is assumed to be initially negligible) near $T \sim m_{\phi}$. The majoron subsequently decays back into neutrinos when it becomes non-relativistic ($T \sim m_{\phi}/3$), generating $\Delta N_{\rm eff } \sim 0.11$.}. Effectively, the model transitions smoothly from two relativistic species to one heated relativistic species. Together this leads to an increase in the contribution to the effective number of neutrino species that occurs sharply around a transition redshift, see also fig. 1 of \cite{Aloni:2021eaq}.

The main factor determining at which redshift such a transition occurs is the mass $m_\chi$\,. Modes that have entered the Hubble horizon far before the transition will have experienced the presence of two relativistic species, while modes that have entered the Hubble horizon far after the transition will only have experienced a single heated relativistic species. If the mass is adjusted precisely such that the time of transition is about a decade of scale factor before recombination, then the CMB multipoles $\ell$ corresponding to the modes that entered the Hubble horizon during the transition will be around $\ell \sim 1000$, allowing the model to more effectively resolve the Hubble tension than either dark species could by itself.

\noindent The general requirements for such a WZDR model are
\begin{enumerate}
    \item a massless strongly self-interacting dark species $\phi$ \label{item:massless}
    \item a massive strongly self-interacting dark species $\xi$\label{item:massive}
    \item a strong interaction between the two species, also allowing for efficient decay of the massive into the massless species\label{item:decay}
    \item no strong interactions with the standard model\label{item:standard}
\end{enumerate}
These requirements allow us to conclude by \cref{item:massless,item:massive} that both species are in local thermal equilibrium, allowing us to write their phase-space distributions as 
\begin{equation}
    f_i(E,t) = \frac{1}{\exp\left(\frac{E-\mu_i(t)}{T_i(t)}+\varepsilon_i\right)}~,\label{eq:distribution}
\end{equation}
with the spin-statistics factor $\varepsilon$ for each species $i$ that can take any of the values $\{-1,0,1\}$. 

While in principle the chemical potentials of either species could be relevant, we will consider here the well motivated\footnote{Usually the involved vertices of interactions will enable interactions with final-state or initial-state \enquote{radiation} (in this case WZDR particles). Such number-changing interactions naturally suppress the chemical potentials, creating chemical equilibrium. A more detailed analysis with a parameter characterizing the strength of these processes involving particle creation or annihilation is left for future work.} case of $\mu_\xi=\mu_\phi=0$. Additionally, the strong interaction of \cref{item:decay} equalizes the two species temperatures into a common thermal bath $T_\xi = T_\phi = T$\,.

\noindent For the massless species, these facts already allows us to conclude that 
\begin{equation}
    3P_\phi = \rho_\phi = s_\phi g_\phi \frac{\pi^2}{30} T_\phi^4~,
\end{equation} 
where $s_\phi$ is a spin-statistics factor, which equals $1$ for bosons, $7/8$ for fermions, and $90/\pi^4$ for a Maxwell-Boltzmann distributed species. Additionally, $g_\phi$ is the number of internal degrees of freedom. Instead, for the massive species we will use the components of the momentum tensor derived from first principles, giving us
\begin{align}
    \rho_\xi &= \frac{g_\xi}{(2\pi)^3} \int p^2 E f_\xi(E,t) \mathrm{d}p~, \label{eq:rho} \\
    3P_\xi &= \frac{g_\xi}{(2\pi)^3} \int p^4/E f_\xi(E,t) \mathrm{d}p~,\label{eq:P}\\
    3 R_\xi &= \frac{g_\xi}{(2\pi)^3} \int E^3 f_\xi(E,t) \mathrm{d}p~,  \label{eq:R}
\end{align}
where we introduced the pseudo-density $R_\xi$ that is important in deriving the equation of motion for the remaining degree of freedom, the temperature $T_\xi=T_\phi = T$\,. The evolution equation for the temperature can be found from the energy density conservation equation\footnote{Which is valid separately for the dark sector due to \cref{item:standard}, completing our four requirements.} (entropy is conserved independently of the precise $T(t)$ evolution). One finds
\begin{equation}
    \frac{\mathrm{d}\ln T_\xi}{\mathrm{d} \ln a} = - \frac{\rho+P}{\rho+R}~,\label{eq:Tevolv}
\end{equation}
where $R=R_\xi + R_\phi = R_\xi + P_\phi$ and $R_\xi$ is given in \cref{eq:R}. The derivation of these statements and more details are given in \cref{app:proof_R}.

To conclude, at the level of the background evolution in this model one need only track a single degree of freedom $T(t)$, which can be used to determine all of \cref{eq:rho,eq:P,eq:R}. Its evolution equation, on the other hand, is given by \cref{eq:Tevolv}. The authors of \cite{Aloni:2021eaq} at this point use an explicit assumption of a Maxwell-Boltzmann distribution of the $\xi$ species, simplifying many of the integrals in terms of Bessel functions. In our case, we will remain agnostic to this assumption and later show that it is indeed well justified in the present context.

At the level of the perturbations, due to the strong interactions assumed in \cref{item:massive,item:massless} one finds simply the typical fluid equations for the species \cite{Ma:1995ey}, with
\begin{align}
    \delta' &= -(1+w) (\theta+\mathcal{M}_c) - 3(c_s^2-w)\mathcal{H} \delta~,\label{eq:delta_dr}\\
    \theta' &= -(1-3 c_s^2) \mathcal{H} \theta + k^2 \frac{c_s^2}{1+w} \delta + \mathcal{M}_e~ \label{eq:theta_dr},
\end{align}
where we used the usual equation of state $w = P/\rho$ and adiabatic sound speed $c_s^2$ (see \cref{app:eq:cs2}). Additionally, we used the continuity metric $\mathcal{M}_c = h'/2$ (synchronous gauge) or $\mathcal{M}_c = -3\phi'$ (conformal Newtonian gauge) as well as the Euler metric $\mathcal{M}_e = 0$ (synchronous gauge) or $\mathcal{M}_e = k^2 \psi$ (conformal Newtonian gauge).

While an instantaneous transition would always have $w = c_s^2 = 1/3$ before and after the $\xi$ species becomes non-relativistic, the true evolution involves regimes in which $w \neq c_s^2$ and where both depart from $1/3$, allowing for additional damping/enhancement of the oscillations inherited by this equation of motion. As such, this model is not purely equivalent to a transition between two self-interacting dark radiation species.

\noindent Lastly, we introduce a few short bits of relevant notation. The ratio of the degrees of freedom will be called $r_g = \frac{g_\xi}{s_\phi g_\phi}$, which in the original supersymmetry-motivated WZDR model is fixed to $8/7$\,. The contribution of this model to the effective number of neutrino species $N_\mathrm{eff}$ is given as
\begin{equation}
    \Delta N(z) = \frac{8}{7} \frac{\rho(z)}{\rho_\gamma(z)} (11/4)^{4/3}~.
\end{equation}
The value this function asymptotes to for $z\to 0$ will be called $N_\mathrm{wzdr}$ for the remainder of this manuscript, while the theoretical asymptotic value as $z\to \infty$ (ignoring the generation of such a species only before or after BBN) is naturally given by $N_\mathrm{wzdr}/(1+r_g)^{1/3}$. We also introduce the approximate scale factor $a_t$ (or redshift $z_t$) of the transition defined as the point at which
\begin{equation}
    (1+z_t) T(z=0) = T(z=0) a_t^{-1} = m_\xi~.\label{eq:zt}
\end{equation}
This is the time when the naive expectation from the decoupled evolution of $\phi$ would reach the mass threshold of $\xi$, signaling the breakdown of this approximate evolution. 

\begin{figure}[t]
    \centering
    \includegraphics[width=0.45\textwidth]{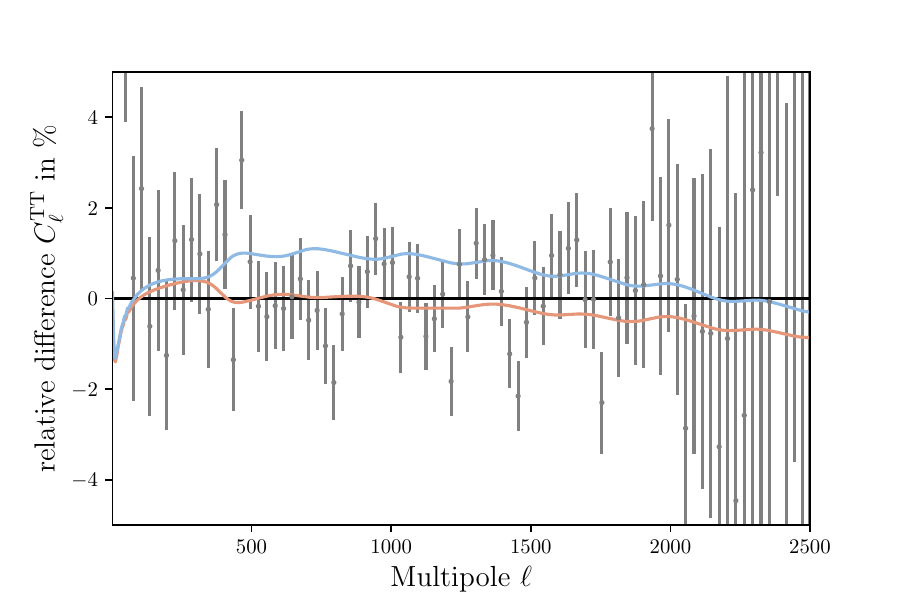}
    \includegraphics[width=0.45\textwidth]{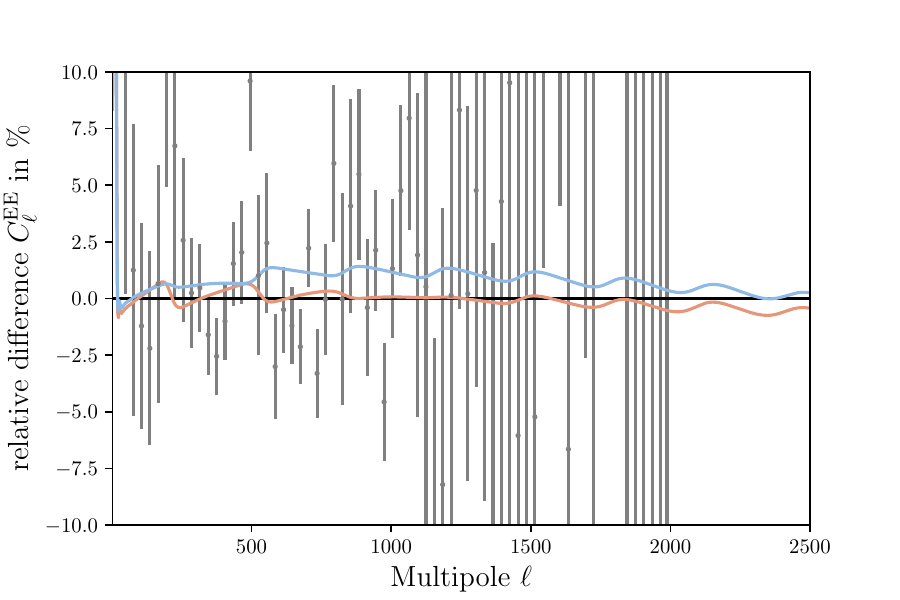}
    \includegraphics[width=0.45\textwidth]{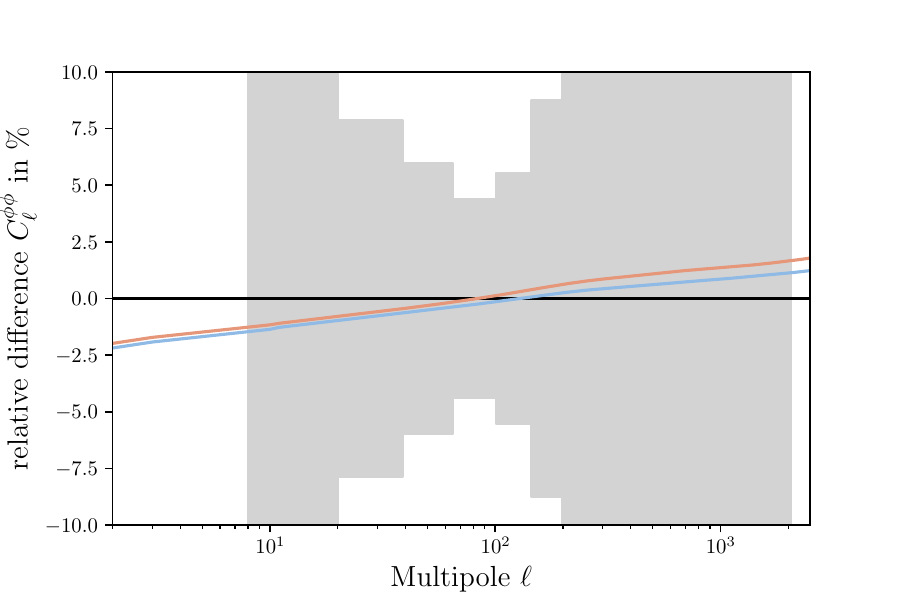}
    \includegraphics[width=0.45\textwidth]{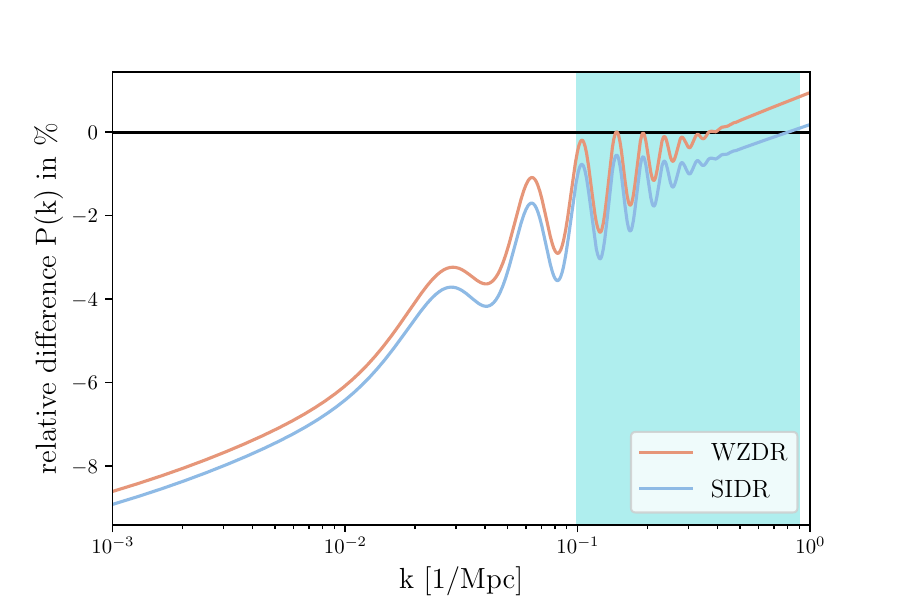}
    \caption{Impact of the WZDR model on various observables, compared to that of self-interacting dark radiation (of equal amount $N_\mathrm{wzdr}=0.3$). The residuals of the models and the data points are taken with respect to the bestfit $\Lambda$CDM model of \cite{Planck:2018vyg}. For the WZDR model we used the bestfit parameters from  a Planck+BAO+Pantheon run with $N_\mathrm{wzdr}=0.3$ fixed, while for the SIDR model we disable the transition. \textbf{Top Left:} TT CMB power spectrum. \textbf{Top Right:} EE CMB power spectrum. \textbf{Bottom Left:} $\phi \phi$ CMB lensing power spectrum. \textbf{Bottom Right:} Matter power spectrum (in blue are the scales relevant for $\sigma_8$).}
    \label{fig:impact_cosmology}
\end{figure}

\noindent We show the impact on cosmological observables of the WZDR model in \cref{fig:impact_cosmology}, where we also show a self-interacting dark radiation model (SIDR) as comparison. We observe that the WZDR model nicely tracks the residuals around $\ell \sim 300$ with a slightly increased suppression at $\ell \gtrsim 1500$, which results in a slightly better fit to CMB data.

The corresponding code is expected to be made public upon acceptance of this manuscript. Additionally, the code can be made available upon request.


\section{WZDR in light of cosmological data}\label{sec:results}

In order to investigate the WZDR model we make use of various commonly used likelihoods, listed below. Our main analysis will focus on the combination of Planck+BAO data, and we will explicitly point out cases where we add additional data.
\pagebreak[20]
\begin{itemize}
    \item \textbf{Planck} -- Planck 2018 (TT + TE + EE + lensing) likelihoods as described in \cite{Planck:2019nip}.
    \item \textbf{BAO} -- Baryonic Acoustic oscillation data from the 6degree of field (6dF) survey \cite{Beutler:2011hx}, the main galaxy sample (MGS) \cite{Ross:2014qpa}, and luminous red galaxies (LRG) \cite{BOSS:2016wmc}.
    \item \textbf{Pantheon} -- Supernovae of Type Ia based on the Pantheon catalog \cite{Pan-STARRS1:2017jku}.
    \item \textbf{ACT} -- CMB data from the Atacama Cosmology Telescope Data Release 4 \cite{ACT:2020gnv}.
    \item \textbf{SPT} -- South Pole Telescope 3rd generation \cite{SPT-3G:2021eoc} and lensing \cite{Wu:2019hek} data.
    \item \textbf{lyaBAO} -- Higher redshift BAO data from eBOSS based on quasars (QSO) \cite{Ata:2017dya} and Lyman-$\alpha$ forest correlations \cite{Blomqvist:2019rah,deSainteAgathe:2019voe}.
    \item \textbf{BBN} -- Measurements of BBN from Cooke et al. \cite{Cooke:2017cwo} and Aver et al. \cite{Aver:2015iza}.
\end{itemize}
Hereafter, in our figures the grey bands indicate the $H_0$ measurement by the SH0ES collaboration \cite{Riess:2021jrx}, while the light-blue bands indicate the combined results on $S_8$ from the combined analysis of KiDS-1000+BOSS+2dFLens from \cite{Heymans:2020gsg}.

\enlargethispage*{1\baselineskip}
Below we examine the assumptions made in \cite{Aloni:2021eaq} for various different aspects of the model, various combinations of data to constrain the model, and the ability to ease the $S_8$ tension when allowing for interactions between cold dark matter and WZDR.

For the purpose of a quick comparison, we will also display the Gaussian tension (GT) values for Hubble parameter. Similarly, we could also compute the $Q_\mathrm{DMAP}$ and $\Delta \mathrm{AIC}$ criteria of \cite{Schoneberg:2021qvd}. However, for the base WZDR model this has already been done in \cite{Aloni:2021eaq}. Instead, we focus on the relative differences compared to this baseline analysis, and for this comparison the GT values are sufficient.

\enlargethispage*{2\baselineskip}
\subsection{Model-based assumptions}
\begin{figure}[t]
    \centering
    \includegraphics[width=0.45\textwidth]{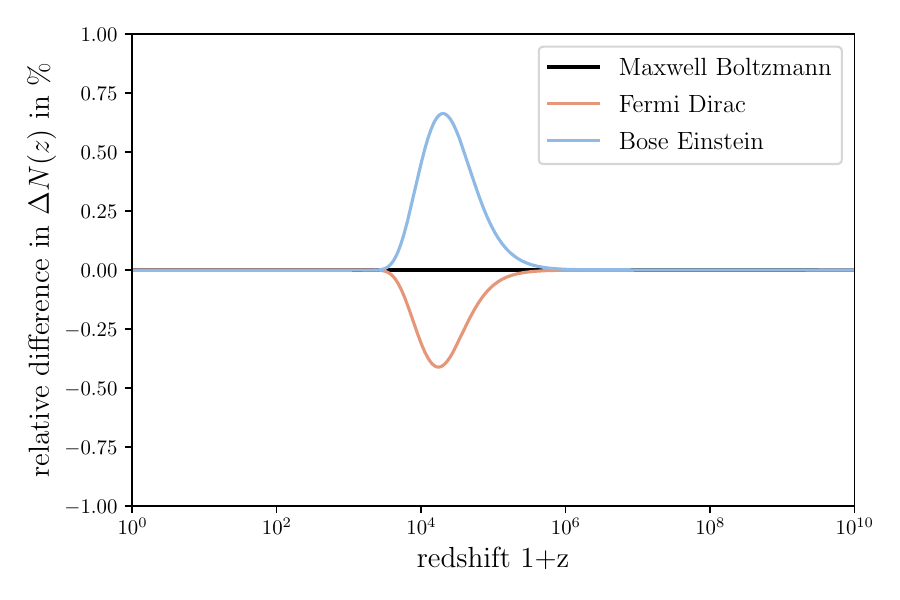}
    \includegraphics[width=0.45\textwidth]{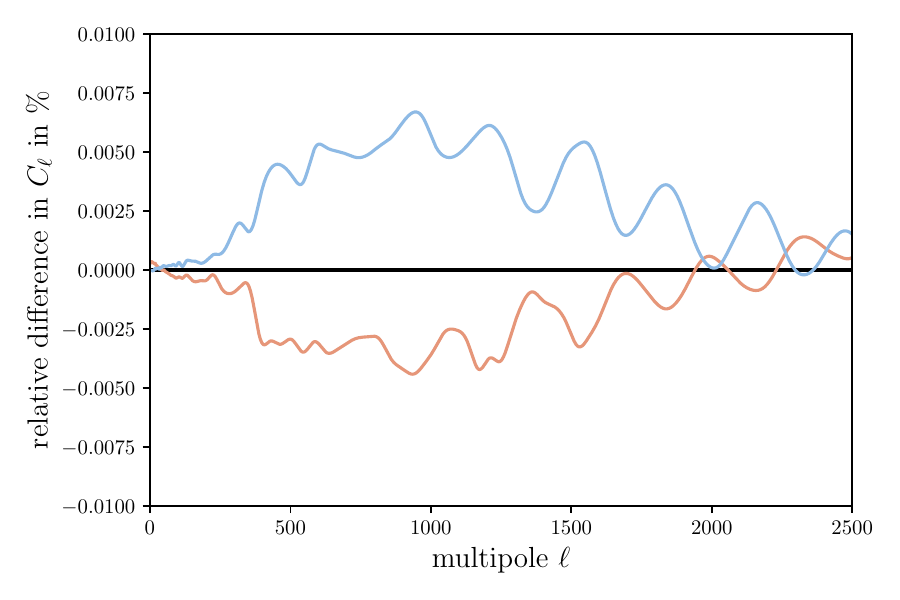}
    \caption{Difference between Maxwell-Boltzmann distributed, Fermi-Dirac distributed, and Bose-Einstein distributed massive species. \textbf{Left:} there is a small but nonzero difference in the number of extra neutrino species from WZDR that does depend on the distribution of the species as it becomes nonrelativistic. \textbf{Right:} The difference in the final angular power spectra of the CMB (here we show the TT autocorrelation) caused by this effect is negligibly small, staying below the sub-permille level. It is comparable to the difference caused by the choice of recombination code. }
    \label{fig:diff_spinstat}
\end{figure}

\begin{figure}[t]
    \centering
    \includegraphics[width=0.45\textwidth]{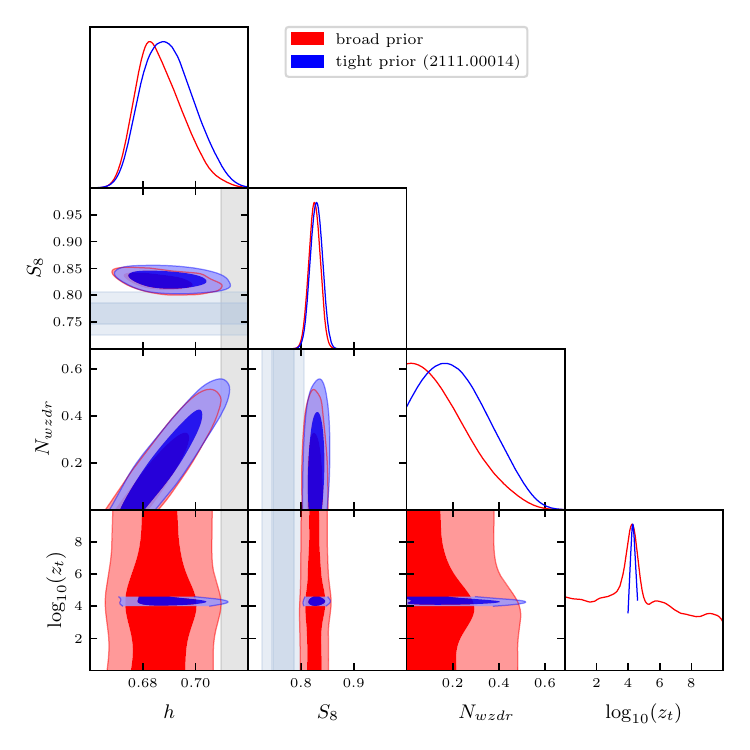}
    \includegraphics[width=0.45\textwidth]{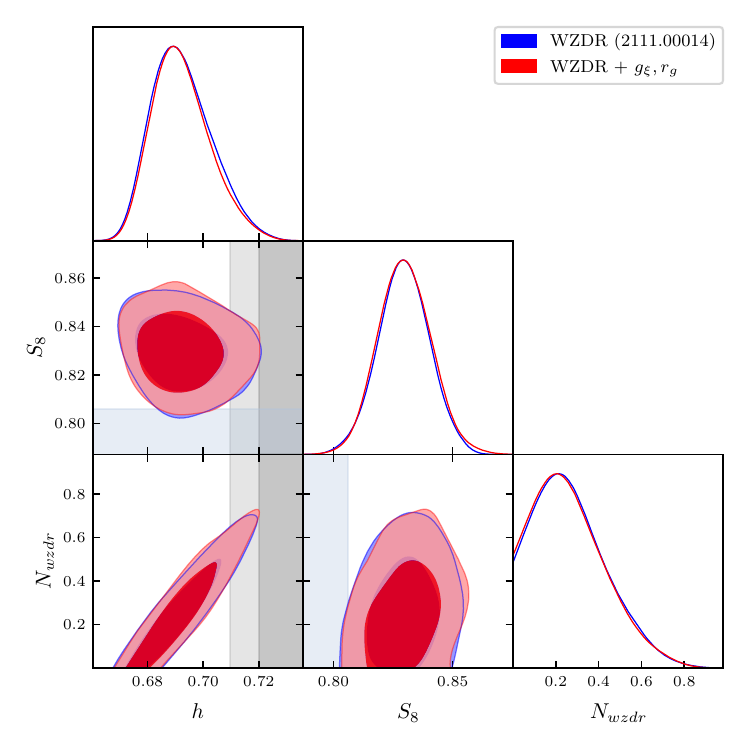}
    \caption{Difference in assumptions of the underlying parameter space of the WZDR model. \textbf{Left:} Difference between tight and broad priorl. The broad prior permissively allows $\log_{10}(z_t)=[0,10]$, while the tighter prior used in \cite{Aloni:2021eaq} restricts it to $\log_{10}(z_t)=[4.0,4.6]$. The tight constraints barely shift the overall contours in $h$ and $N_\mathrm{wzdr}$ and appear as a natural extension of the broad prior constraints. \textbf{Right:} Difference between the default analysis (with a tight prior) and one where the parameters $r_g$ and $g_\xi$ are additionally varied. See the text for an explanation on the bands.
    } 
    \label{fig:diff_prior}
\end{figure}

\begin{figure}[t]
    \centering
    \includegraphics[width=0.47\textwidth]{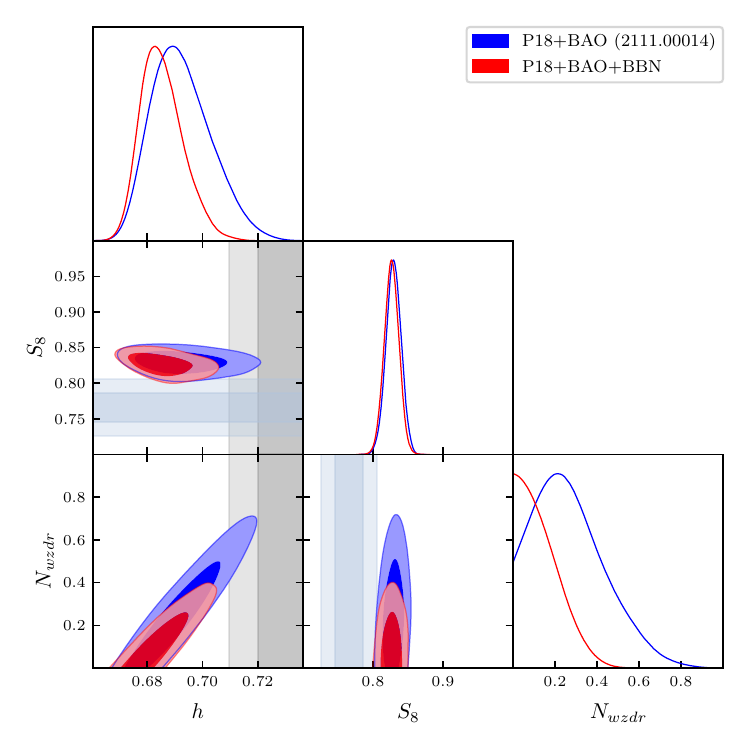}
    \includegraphics[width=0.43\textwidth]{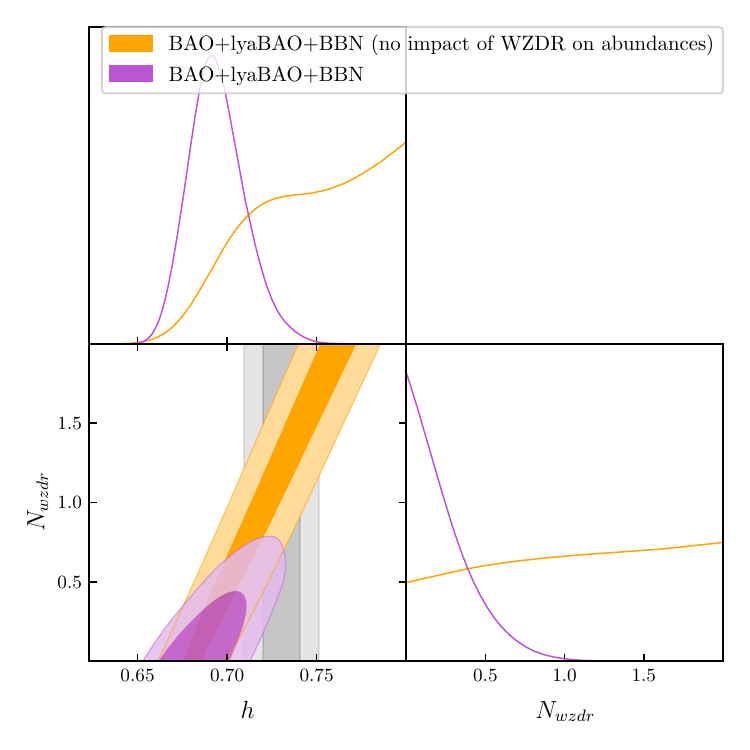}
    \caption{Comparison of the constraints on $H_0$\,, $S_8$\,, and $N_\mathrm{wzdr}$ when BBN data are considered. \textbf{Left:} Comparison of the standard analysis with and without considering BBN data (assuming an impact of WZDR on the abundances of light elements). \textbf{Right:} Comparison of the BAO+BBN probe between assuming an impact of WZDR on the abundances of light elements and assuming no impact.
    }
    \label{fig:diff_bbn}
\end{figure}

Given the excellent ability of the model of \cite{Aloni:2021eaq} to ease the Hubble tension, we check here what kinds of assumptions are beneficial to this ability. We are especially interested in the robustness of the resolution of the Hubble tension for different analysis choices of \cite{Aloni:2021eaq}. As we describe below the model remains almost equally as able to ease the Hubble tension even through a variety of possible model choices, or the choice is at otherwise very well motivated.
\begin{itemize}
    \item Maxwell-Boltzmann distribution: This assumption has virtually no impact on the posteriors, as expected since the predicted CMB angular power spectra are almost identical, as we show in \cref{fig:diff_spinstat}. This also confirms the assertion of \cite{Aloni:2021eaq}. Nevertheless, the newly developed formalism also allows to probe phase-space distributions which are non-thermal, as long as the distribution function depends solely on the parameter combination $E/T$.
    \item Width of the prior: Due to the Bayesian nature of the analysis, this choice naturally has a big impact on the derived constraints. If we extend the width of the prior to $\log_{10}(z_t) \in [0,10]$, the Hubble parameter constraints are $H_0 = 68.7^{+0.7}_{-1.1} \mathrm{km/s/Mpc}$ as the Bayesian prior volume where the model reduces mostly to $\Lambda$CDM are driving the constraints, whereas the small volume $\log_{10}(z_t) \in [4,4.6]$ where the model eases the Hubble tension only gives a small contribution. Considering  \cref{fig:diff_prior}, there does not appear a fundamental problem with this choice, however. The minimum in both cases is the same (it lies in the tight region), with a $\chi^2$ around 2780.6. As such, we believe that the restriction of the prior to the relevant region does slightly increase the inferred value of $H_0$ in an Bayesian analysis, but this choice is well justified.
    \item Species abundances: The abundances of both species are fixed through $g_\xi=2$ and $r_g=8/7$ as well as specification of their distribution and temperature. Allowing $r_g$ to vary leads to the more general \enquote{Stepped Dark Radiation} model of \cite{Aloni:2021eaq}, while the variation of $g_\xi$ (the internal degrees of freedom of the $\xi$ species) simply scales the overall abundances of both species for a given temperature. Allowing these two parameters to vary has only a small impact on the posteriors, as seen in \cref{fig:diff_prior}.
    \item Big Bang Nucleosynthesis (BBN) impact: In \cite{Aloni:2021eaq} the authors motivate that the impact of the WZDR species on BBN can be treated as negligible if one assumes that the WZDR density is enhanced only after BBN, such as when created from the decay of a massive particle shortly after BBN. A window for such decays stretches\footnote{The highest wavenumbers relevant for the CMB as measured with Planck (around $k=0.15 h/\mathrm{Mpc}$ \cite{Planck:2018jri}) enter the Hubble horizon ($k=aH$) in radiation domination giving $z \sim 10^{4.8}$ for $N_\mathrm{eff}$ not too far from $3$. Instead, BBN occurs around $0.07 \ \mathrm{MeV}$ (due to the mass threshold of hydrogen fusion and the low abundance of baryons compared to photons), giving $z \sim 10^{8.5}$.} from around $z \sim 10^{4.8}$ to $z \sim 10^{8.5}$, allowing for a reasonable time for a decay to occur. However, this also requires the introduction of additional dark-sector particles or interactions not present in the simplest motivations, and a small degree of fine-tuning. Given the strong constraints of BBN on additional dark radiation (see \cref{fig:diff_bbn}, left panel), it is crucial to introduce such additional motivation if the model is to ease the Hubble tension. When considering an impact on the light element abundances as measured by Cooke et al. \cite{Cooke:2017cwo} and Aver et al.  \cite{Aver:2015iza}, the constraint on $h$ increases to $H_0=68.54^{+0.60}_{-0.85} \mathrm{km/s/Mpc}$ due to these BBN measurements,  excluding the model as a viable resolution the Hubble tension.
\end{itemize}

\begin{figure}[t]
    \centering
    \includegraphics[width=0.45\textwidth]{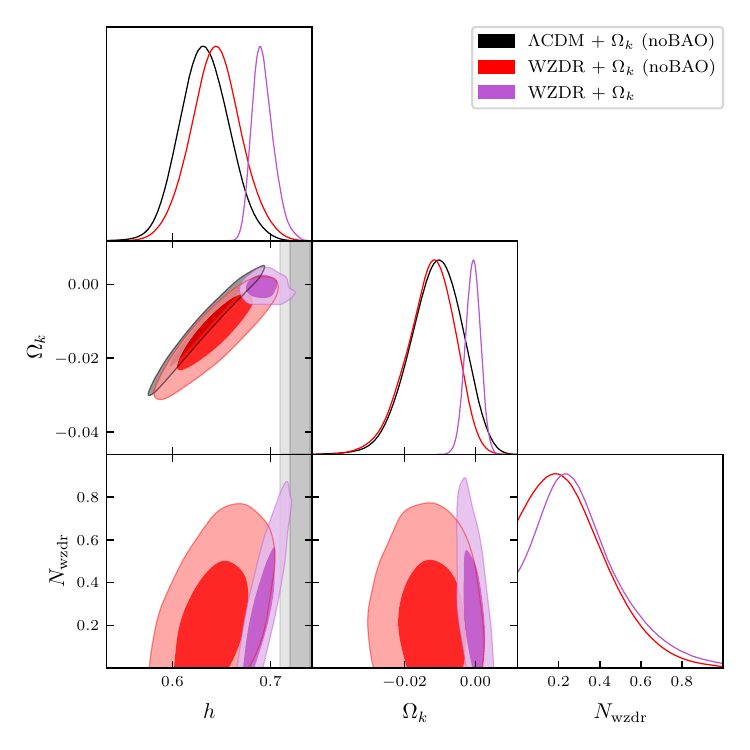}
    \includegraphics[width=0.45\textwidth]{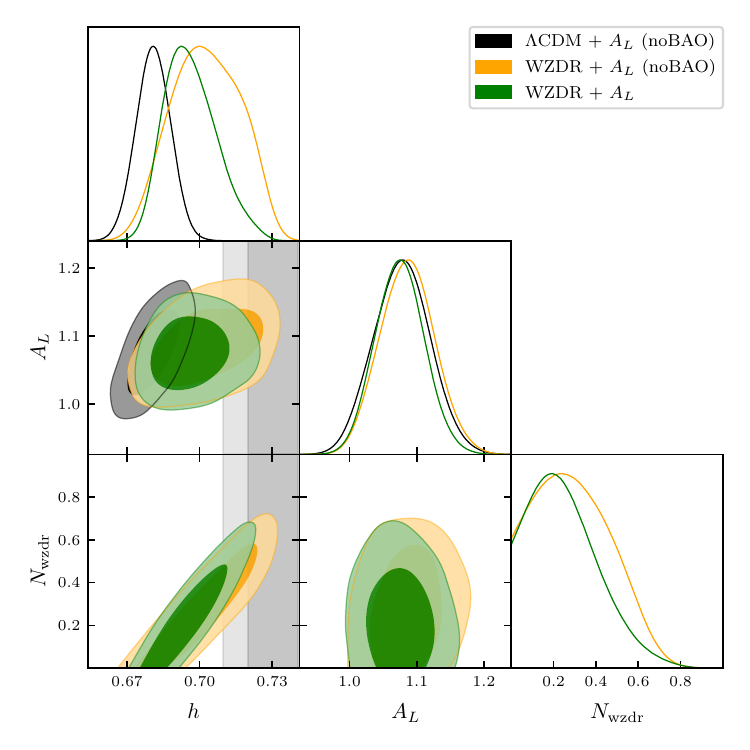}
    \caption{Comparison of the constraints on $H_0$\, $N_\mathrm{wzdr}$, and the extra parameter for two one-parameter-extensions (with and without BAO data included). \textbf{Left:} Extension with $\Omega_k$\,. \textbf{Right:} Extension with $A_L$\,.}
    \label{fig:okal}
\end{figure}

Another natural question is whether the easing of the Hubble tension also results in other existing CMB tensions in the context of Planck data being eased as well. In particular, we investigate whether for a WZDR model there is still preference for non-zero $\Omega_k$ and $A_L$ different from unity. In \cref{fig:okal} we observe that neither extension by $\Omega_k$ or $A_L$ allow for significantly different $H_0$ values (we remind the base data include BAO and CMB lensing).

Furthermore, it is visible that neither of the tensions is eased by a great amount. Indeed, the constraint $\Omega_k \sim 0$ is consistently recovered when considering BAO data ($10^3\Omega_k = 1.27^{+2.09}_{-2.18}$) and a slight preference is present for $\Omega_k < 0$ in the absence of BAO data ($10^3\Omega_k = -12.7^{+9.1}_{-5.3}$), as in the case for $\Lambda$CDM.\enlargethispage*{2\baselineskip}
 We also observe that $A_L$ is even a slight bit more in tension with $A_L=1$ ($A_L = 1.089\pm0.042$, GT $2.1\sigma$) compared to $\Lambda$CDM ($A_L = 1.079 \pm 0.042$, GT $1.9\sigma$).

\begin{figure}[t]
    \centering
    \includegraphics[width=0.45\textwidth]{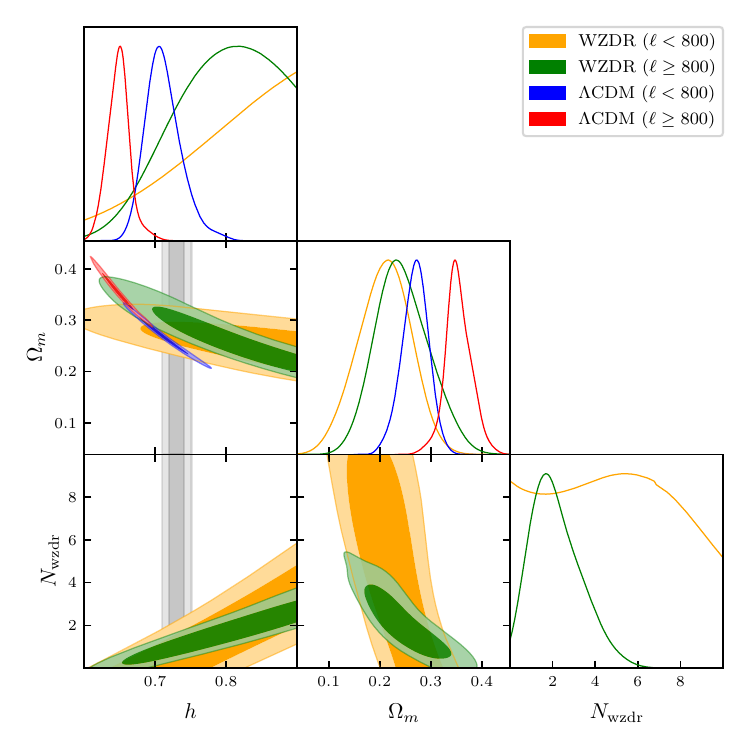}
    \caption{Comparison of the constraints on $H_0$\, $N_\mathrm{wzdr}$, and $\Omega_m$ from the CMB temperature autocorrelation at $\ell < 800$ and $\ell \geq 800$. We show both $\Lambda$CDM and WZDR.}
    \label{fig:l800}
\end{figure}

As for the difference between the constraints from temperature autocorrelation between $\ell < 800$ and $\ell \geq 800$ \cite{Planck:2016tof}, we observe in \cref{fig:l800} that the WZDR model completely eliminates the tension that is clearly visible in the $\Lambda$CDM model. One finds in $\Lambda$CDM the constraints on the matter density of $\Omega_m = 0.268 \pm 0.028$ for $\ell < 800$ and $\Omega_m = 0.356 \pm 0.039$ for $\ell \geq 800$, a $1.8\sigma$ tension. Similarly, in the WZDR model we find $\Omega_m = 0.208 \pm 0.053$ for $\ell < 800$ and $\Omega_m = 0.244 \pm 0.055$ for $\ell \geq 800$, only a $0.5\sigma$ tension. However, we note that the tension is mostly eased towards very high $h$ and $N_\mathrm{wzdr}$ values which are disfavored by polarization measurements. Still, we conclude that (as expected) this particular tension is eased in the WZDR model due to its ability to differently affect scales at $\ell < 800$ and $\ell \geq 800$.

All results for testing different model assumptions are summarized in  \cref{tab:model_results}.

\begin{table}[t]
    \centering
    \renewcommand{\arraystretch}{1.5}
    \begin{tabular}{c|c c | c c}
        Model & $H_0$ [km/s/Mpc] & GT & $S_8$ & $N_\mathrm{wzdr}$\\ \hline
        Default assumptions & $69.2^{+1.0}_{-1.3}$ & 2.7$\sigma$ & $0.829 \pm 0.011$ & $<0.605$ \\
        Wider prior & $68.7^{+0.7}_{-1.1}$ & 3.5$\sigma$ & $0.826 \pm 0.011$ & $<0.453$\\
        Including BBN & $68.5^{+0.6}_{-0.9}$ & $3.8\sigma$ &  $0.826 \pm 0.011$ & $<0.325$\\
        Varying $r_g\,, g_\xi$ & $69.2^{+0.9}_{-1.2}$ & 2.8$\sigma$ & $0.830 \pm 0.011$ & $<0.590$\\
        Including $A_L$ & $69.7^{+0.9}_{-1.3}$ & 2.4$\sigma$ & $0.809 \pm 0.011$ & $<0.571$\\
        Including $\Omega_k$ & $69.5^{+1.0}_{-1.4}$ & 2.5$\sigma$ & $0.827 \pm 0.012$ & $<0.803$\\
    \end{tabular}
    \caption{Comparison of the results for the variations of model assumptions. The Gaussian Tension (GT) is considered with respect to the analysis of \cite{Riess:2021jrx}.}
    \label{tab:model_results}
\end{table}

\subsection{Data-based assumptions}
\begin{figure}[t]
    \centering
    \includegraphics{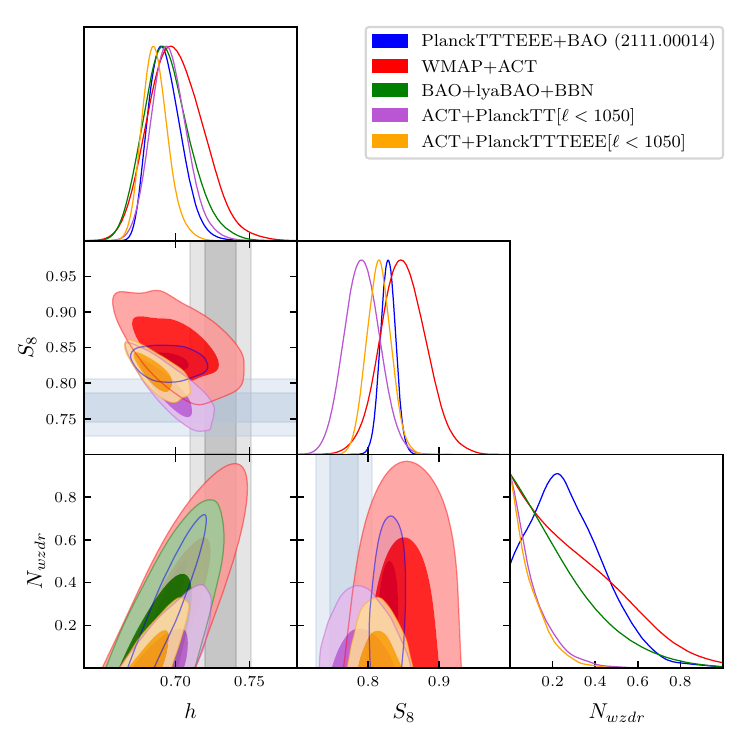}
    \caption{Comparison of the constraints on $H_0$\,, $S_8$\,, and $N_\mathrm{wzdr}$ for various data sets beyond Planck. In blue we show results corresponding to the analysis of \cite{Aloni:2021eaq}. The bands are shown according to the introduction to \cref{sec:results}.}
    \label{fig:data_results}
\end{figure}

\begin{figure}[t]
    \centering
    \includegraphics[width=0.8\textwidth]{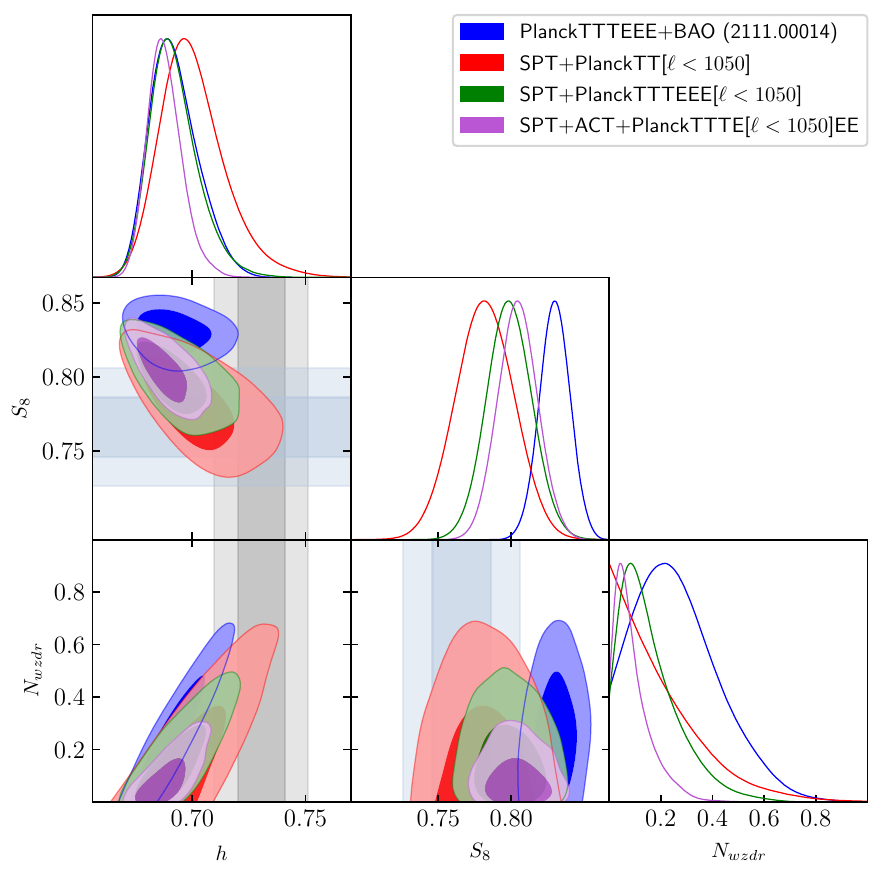}
    \caption{Same as \cref{fig:data_results} but with data combinations including SPT.}
    \label{fig:data_results2}
\end{figure}

Not only the choices related to the model itself need to be checked in such an analysis, also the data used. We extend the original analysis with alternative data sets to those chosen in \cite{Aloni:2021eaq}, focusing on alternative data combinations with similar constraining power to Planck data. We consider for the combinations with ACT and SPT (see the beginning of \cref{sec:results} for definitions) only non-overlapping CMB data (except in our most constraining run with ACT+SPT).
The results are summarized in \cref{tab:data_results,fig:data_results,fig:data_results2}.

\begin{table}[t]
    \centering
    \renewcommand{\arraystretch}{1.5}
    \begin{tabular}{c|c c | c c}
        Data & $H_0$ [km/s/Mpc] & GT & $S_8$ & $N_\mathrm{wzdr}$\\ \hline
        Planck+BAO & $69.2^{+1.0}_{-1.3}$ & 2.7$\sigma$ & $0.829 \pm 0.011$ & $<0.605$\\
        BAO+lyaBAO+BBN & $69.4^{+1.4}_{-1.9}$ & 2.1$\sigma$ & Unconstrained & $<0.628$ \\
        ACT+WMAP & $69.8^{+1.4}_{-2.2}$ & 1.9$\sigma$ & $0.849 \pm 0.032$  & $<0.725$\\
        ACT+PlanckTT[$\ell<1050$] & $69.6^{+1.1}_{-1.3}$ & 2.3$\sigma$ & $0.792 \pm 0.025$ & $<0.292$\\
        ACT+P1anckTTTEEE[$\ell<1050$] & $68.6^{+0.77}_{-1.0}$ & 3.4$\sigma$ & $0.815 \pm 0.017$ & $<0.254$\\
        SPT+PlanckTT[$\ell<1050$]& $70.0^{+1.1}_{-1.6} $ & 2.0$\sigma$ & $0.782 \pm 0.02$  & $<0.547$\\
        SPT+PlanckTTTEEE[$\ell<1050$] & $69.2^{+0.8}_{-1.3}  $ & 2.9$\sigma$ & $0.799 \pm 0.016$ & $<0.392$ \\
        SPT+ACT+PlanckTTTE[$\ell<1050$]EE &  $68.8^{+0.6}_{-0.9}$ & 3.5$\sigma$ & $0.804 \pm 0.014$ & $<0.239$
    \end{tabular}
    \caption{Results for the variations of data. The Gaussian Tension (GT) is considered with respect to the analysis of \cite{Riess:2021jrx}.}
    \label{tab:data_results}
\end{table}

\newpage

\noindent We consider the following combinations of data sets:
\begin{itemize}
    \item BAO + lyaBAO + BBN : We combine our usual probes of BAO with those from higher redshifts (which we refer to as lyaBAO) as well as the measurements of BBN from Cooke et al. \cite{Cooke:2017cwo} and Aver et al. \cite{Aver:2015iza} in order to probe the model completely independently from any CMB data, see also \cite{Schoneberg:2019wmt} for the equivalent analysis in the $\Lambda$CDM model. 
    In \cite{Schoneberg:2019wmt} it has been shown that such a combination is very sensitive to solutions to the Hubble tension involving additional dark radiation. We observe in \cref{fig:data_results} that this combination indeed strongly constrains the Hubble tension, leading to $H_0 = 69.2^{+1.0}_{-1.3} \mathrm{km/s/Mpc}$\, ($2.1\sigma$ GT). Naturally, this is based on the assumption that the abundances of the species $\phi$ and $\xi$ are already present during BBN, and lead to measurably different light element abundances due to their impact on the expansion history. However, if one assumes these species to be generated after BBN has already finished, then naturally we do not recover any constraint on $H_0$ with this probe, as $N_\mathrm{wzdr}$ becomes completely unconstrained (\cref{fig:diff_bbn}, right panel).
    
    \item ACT + WMAP : We combine data from WMAP with those from the ACT DR 4. This probe covers both small and large multipoles $\ell$ similar to Planck, albeit with a reduced sensitivity especially at intermediate multipoles. As such, we observe weaker bounds on $N_\mathrm{wzdr}$ from this combination (the weakest bounds of all combinations), leading to a rather permissive $H_0 = 69.8^{+1.4}_{-2.2} \mathrm{km/s/Mpc}$ ($1.9\sigma$ GT). We thus observe that the intermediate multipoles (especially from polarization, see below) are fundamental in constraining the WZDR model. However, it is worth pointing out that we observe weaker bounds, but not a detection of $N_\mathrm{wzdr}$ (unlike for EDE \cite{Poulin:2021bjr,Hill:2021yec,Smith:2022hwi}). As such, at present there is not a strong hint of a presence of WZDR yet. 
    
    \item ACT + PlanckTT[$\ell < 1050$]  : We can also trade the WMAP data for Planck data, creating a probe with different systematics dependence than the full Planck data. We consider Planck data only below $\ell \sim 1050$, which is approximately the range of WMAP data, and also where the Planck-inferred parameters begin to drift between smaller and higher multipoles \cite{Planck:2016tof}. In this case we do recover relatively tight constraints on $N_\mathrm{wzdr}$, leading to $H_0 = 69.6^{+1.1}_{-1.3} \mathrm{km/s/Mpc}$ ($2.3 \sigma$ GT). This combinations is only slightly less constraining  than the full Planck data, showing the importance of the intermediate-to-high multipole information ($\ell \sim 1000-2500$) in shifting the mean, as well as that of the polarization.
    
    \item ACT + PlanckTTTEEE[$\ell <1050$] : Indeed, if we instead consider also polarization data for Planck (still at $\ell \lesssim 1050$), we find a much tighter constraint leading to $H_0 = 68.6^{+0.77}_{-1.0} \mathrm{km/s/Mpc}$ ($3.4\sigma$ GT), even strengthening the bound on WZDR compared to Planck data. This is most likely caused by the excellent sensitivity of the TE and EE spectra to additional dark radiation species, especially towards higher $\ell$ which are more accurately probed by ACT data.
    
    \item SPT + PlanckTT[$\ell < 1050$] : The same analysis can be made when considering SPT-3G data instead of ACT data. In this case the preference for high $H_0$ present in SPT-3G data compared to ACT data for $\Lambda$CDM also shows in slightly ($\sim$0.5$\sigma$) higher mean values, while the error bars remain comparable. Thus, the conclusion from ACT data remains roughly the same. When using only the temperature autocorrelation data from Planck we find $H_0 = 70.0^{+1.1}_{-1.6} \mathrm{km/s/Mpc}$. 
    
    We also note that the reconstructed $S_8$ values when including SPT data tend to be in better agreement with the determinations weak-lensing measurements. This is not surprising, since it is known that the $\Lambda$CDM cosmology deduced from SPTPol \cite{SPT:2017jdf} and SPT-3G \cite{SPT-3G:2021eoc} has a smaller $A_s$ and $\Omega_m h^2$, leading to low $S_8$ values (similarly to the $\Lambda$CDM cosmology deduced from Planck when $A_L$ is marginalized over \cite{Motloch:2018pjy,DiValentino:2018gcu}). Hence, when including only PlanckTT[$\ell <1050$], we reconstruct $S_8 = 0.782 \pm 0.020$.
    
    \item SPT + PlanckTTTEEE[$\ell <1050$] : When we also add polarization data for Planck (still at $\ell \lesssim 1050$), the constraint on $H_0$ tightens to $H_0 = 69.2^{+0.8}_{-1.3} \mathrm{km/s/Mpc}$. This value is slightly larger than the one inferred in the  ACT + PlanckTTTEEE[$\ell <1050$] analysis, consistent with the fact that SPT-3G data generally prefers higher mean values of $H_0$ than ACT data. On the other hand, the addition of polarization data drives $S_8$ to a higher value $S_8 = 0.799 \pm 0.016$, which lies in between the weak-lensing and the Planck-only determinations.
    
    \item ACT + SPT + PlanckTTTE[$\ell <1050$]EE : It was shown in \cite{Smith:2022hwi} that the combination of ACT DR4, SPT-3G, Planck polarization and low-medium $\ell$ Planck temperature data led to a $3.3\sigma$ preference for EDE,
   highlighting the important role played by the high $\ell$ Planck polarization in setting this preference. We have carried an analogous analysis for the WZDR model, in order to determine whether a similar preference could arise for $N_{\rm wzdr}$. We found that the WZDR component is not detected with this data combination, obtaining instead the strongest constraint on $N_{\rm wzdr} < 0.239$, and a corresponding low value of the Hubble parameter, $H_0 = 68.8^{+0.6}_{-0.9} \mathrm{km/s/Mpc}$ ($3.5\sigma$ GT). This confirms that the accurate polarization measurements from the different CMB datasets severely limit any addition of dark radiation, even in the more flexible setting of the WZDR model.
\end{itemize}

\noindent We conclude that the CMB data constrain the WZDR model especially through higher multipole $\ell$ polarization measurements. Releasing this polarization accuracy (even at low-intermediate scales, such as with WMAP) causes looser constraints, while using data with higher accuracy causes tighter constraints. None of the existing data show a clear preference for the WZDR model. Considering an impact on BBN only further constraints the model, leading to excellent bounds from the combination of BAO+BBN even independently from CMB data.

\subsection{Adding interacting DM: implications for the $S_8$ tension}
\label{sec:results_S8}

The prediction for $S_8$ in the WZDR model is not significantly altered with respect to $\Lambda$CDM (see \cref{tab:data_results}), so it is legitimate to ask whether the addition of interactions between the WZDR species and the cold dark matter (DM) could possibly also aid in reducing the $S_8$ tension. As already pointed out in \cite{Aloni:2021eaq}, it is very natural to expect a coupling between a dark matter component and the dark radiation. Such coupling equips the dark matter with a pressure supports that reduces the growth of structure, thereby leading to a smaller $S_8$ \cite{Buen-Abad:2015ova,Lesgourgues:2015wza,Buen-Abad:2017gxg, Archidiacono:2019wdp,Becker:2020hzj}. Here we want to check whether this straightforward addition can indeed address both the $H_0$ and $S_8$ tensions successfully. To that end, we use the generic ETHOS parametrisation for interactions between dark radiation (DR) and cold dark matter (DM) \cite{Cyr-Racine:2015ihg}, where the DR-DM scattering rate scales like a power-law of the temperature, $T^n$. 

Since we are considering a self-interacting WZDR fluid, the DR equations \eqref{eq:delta_dr}-\eqref{eq:theta_dr} remain the same except for the addition of a term in the Euler equation accounting for the momentum exchange between the DM and the WZDR
\begin{align}
    \delta' &= -(1+w) (\theta+m_c) - 3(c_s^2-w)\mathcal{H} \delta~,\label{eq:delta_dr_dm}\\
    \theta' &= -(1-3 c_s^2) \mathcal{H} \theta + k^2 \frac{c_s^2}{1+w} \delta + m_e + \Gamma_{\mathrm{DR-DM}} (\theta -\theta_{\rm DM}) \label{eq:theta_dr_dm}~,
\end{align}
The effective comoving scattering rate $\Gamma_{\mathrm{DR-DM}}$ is parametrised in the following way \cite{Archidiacono:2019wdp}
\begin{equation}
\Gamma_{\mathrm{DR-DM}} = -\Omega_\dm h^2 a_\dark \left( \frac{1+z}{1+z_d} \right)^n~, 
\label{eq:scat_rate}
\end{equation}
where $z_d=10^7$ is a normalization factor, the index $n$ gives the temperature dependence and $a_\dark$ denotes the interaction strength. On the other hand, the continuity and Euler equation for the cold DM perturbations read 
\begin{align}
\delta'_\dm &= -\theta_\dm -\mathcal{M}_c~, \label{eq:delta_dm_dr} \\ 
\theta'_\dm &= -\mathcal{H} \theta_\dm + k^2 c^2_\dm \delta_\dm + \mathcal{M}_e + \Gamma_{\mathrm{DM-DR}} (\theta_\dm -\theta)~.
\end{align}
Here $c_\dm$ and $\Gamma_{\mathrm{DM-DR}}$ indicate the dark matter sound speed and the effective scattering rate of DM off WZDR. The latter can be simply obtained from energy-momentum conservation, 
\begin{equation}
\Gamma_{\mathrm{DM-DR}} =\frac{(1+w)\rho}{\rho_\dm}  \Gamma_{\mathrm{DR-DM}}~.
\end{equation}
Note that far away from the step in $\Delta N(z)$ we have $w = 1/3$, so we recover the same relation $\Gamma_{\mathrm{DM-DR}} = \left(\frac{4}{3} \frac{\rho}{\rho_\dm} \right) \Gamma_{\mathrm{DR-DM}}$ as in former works \cite{Buen-Abad:2015ova,Lesgourgues:2015wza,Buen-Abad:2017gxg,Archidiacono:2019wdp, Becker:2020hzj}. The DM sound speed is obtained from the following expression \cite{Becker:2020hzj}
\begin{equation}
c_\dm^2 = \frac{k_b T_\dm}{m_\dm} \left(1-\frac{1}{3} \der{\ln T_\mathrm{DM}}{\ln a} \right)~,
\end{equation}
where $m_\dm$ is the dark matter mass (that we fix\footnote{As long as the DM mass is sufficiently high ($m_{\rm DM} > 1 \ \rm{MeV}$), the choice of DM mass has no impact, since the pressure source term $k^2 c_\dm^2 \delta_\dm$ is always negligible on cosmological scales (see \cite{Becker:2020hzj}), and there is no other direct influence of the dark matter mass in the evolution equations. This non-appearance also relates to the assumption of \textit{cold} dark matter.} to $m_\dm = 100 \ \rm{GeV}$), and the DM temperature evolves according to
\begin{equation}
T'_\dm = -2\mathcal{H}T_\dm -2\Gamma_{\mathrm{DM-DR}} (T_\dm - T)~.
\end{equation}
Regarding the redshift dependence of the scattering rate, we will restrict ourselves to the case $n=0$, which arises in the Non-Abelian Dark Matter (NADM) model of \cite{Buen-Abad:2015ova}. This choice is motivated by the twin facts that the NADM model predict strong self-interactions in the dark sector (just as in the original WZDR model) and that it has been previously shown to alleviate the $S_8$ tension \cite{Buen-Abad:2017gxg,Archidiacono:2019wdp,Schoneberg:2021qvd} \footnote{One should still check whether the DR-DM scattering rate in \eqref{eq:scat_rate} with $n=0$ is compatible with the original particle physics setting of the WZDR model, so this exercise should not be interpreted beyond a purely phenomenological description.}. Hence, the extra free parameters in this framework are $\{N_\mathrm{wzdr} , z_t, a_\dark\}$. To ease the comparison with \cite{Lesgourgues:2015wza, Buen-Abad:2017gxg, Archidiacono:2019wdp}, it is convenient to trade the parameter $a_\dark$ for the current scattering rate of DM off DR, which in the case $n=0$ simply reads
\begin{equation}
\Gamma_0 \equiv \Gamma_{\mathrm{DM-DR}} (z=0) = \frac{4}{3} \Omega_\mathrm{WZDR} h^2 a_\dark = 7.49 \cdot 10^{-6}\cdot N_\mathrm{wzdr} \cdot  a_\mathrm{dark}~. 
\end{equation}

\begin{figure}[t]
    \centering
    \includegraphics[width=0.45\textwidth]{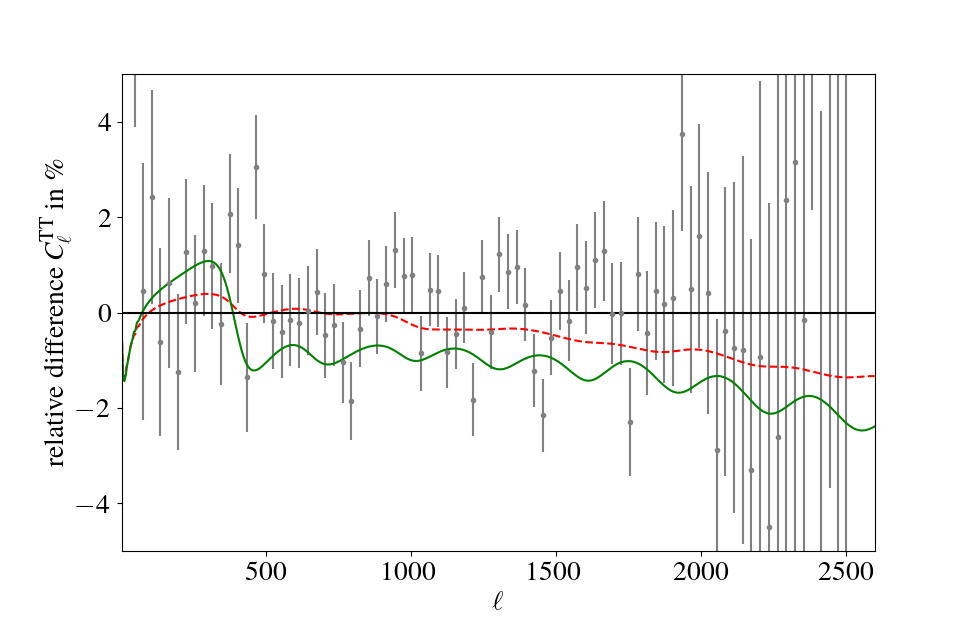}
    \includegraphics[width=0.45\textwidth]{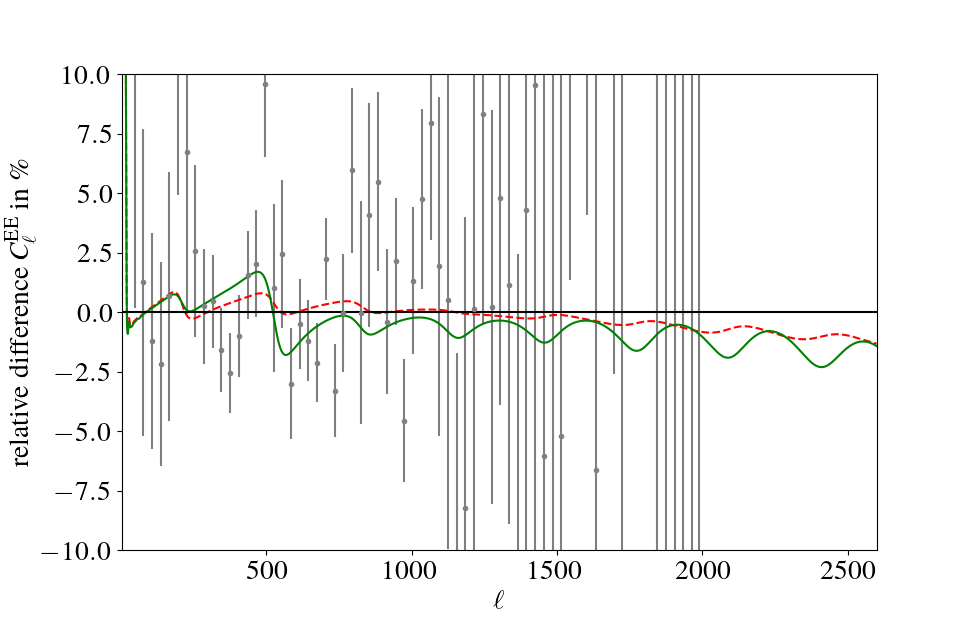}
    \includegraphics[width=0.45\textwidth]{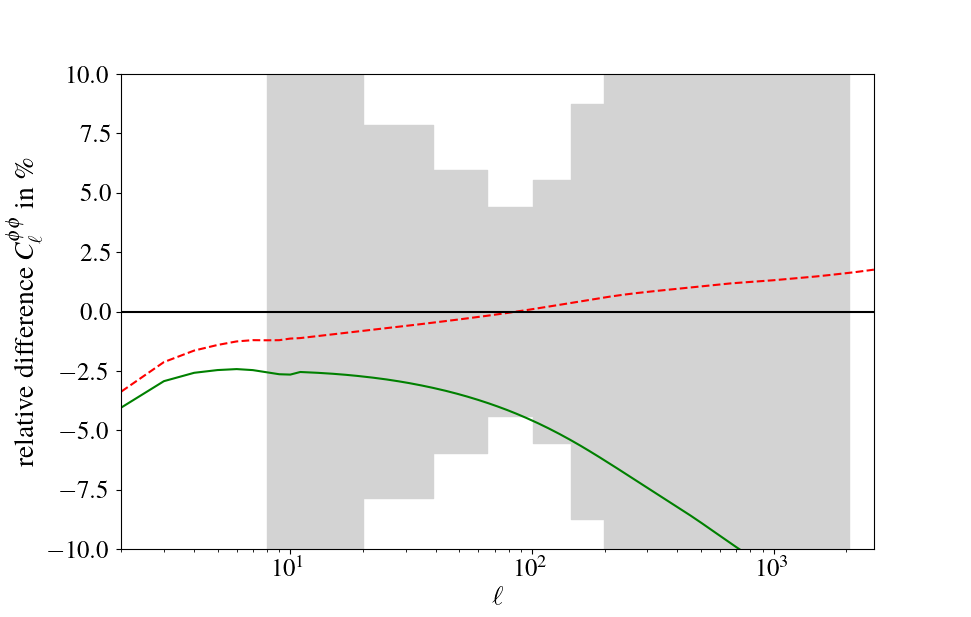}
    \includegraphics[width=0.45\textwidth]{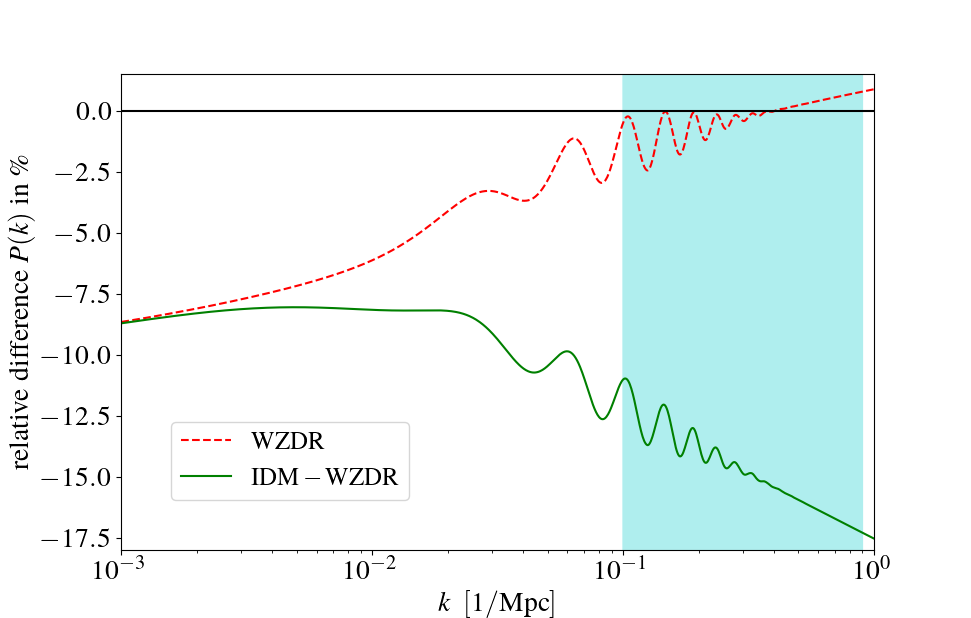}
    \caption{Impact of the WZDR and IDM-WZDR models (of equal amount $N_\mathrm{wzdr}=0.3$) on various observables. The residuals of the models and the data points are taken with respect to the bestfit $\Lambda$CDM model of \cite{Planck:2018vyg}. For the WZDR model we used the bestfit parameters for a Planck+BAO+Pantheon run with $N_\mathrm{wzdr}=0.3$ fixed, while for the IDM-WZDR model we added an interacting rate $\Gamma_0 =10^{-7} \ \mathrm{Mpc}^{-1}$, chosen to give $\sigma_8 \simeq 0.77$, in agreement with weak-lensing measurements. \textbf{Top Left:} TT CMB power spectrum. \textbf{Top Right:} EE CMB power spectrum. \textbf{Bottom Left:} $\phi \phi$ CMB lensing power spectrum. \textbf{Bottom Right:} Matter power spectrum (in blue are the scales relevant for $\sigma_8$).}
    \label{fig:spectra_idm_wzdr}
\end{figure}

In order to illustrate the cosmological impact of adding DM interactions to the WZDR model, we report in the bottom right panel of \cref{fig:spectra_idm_wzdr} the residuals in the present linear matter power spectrum for two different models: 1) the original WZDR scenario with the bestfit parameters for a Planck+BAO+Pantheon run with $N_\mathrm{wzdr}=0.3$ fixed, and 2) the same WZDR model, but including interactions with the totality of the DM (IDM+WZDR) with a current scattering rate $\Gamma_0 = 10^{-7} \ \rm{Mpc}^{-1}$. This value of $\Gamma_0$ is simply chosen because it yields $\sigma_8 \simeq 0.77$, in close agreement with the weak-lensing determinations. We see that the interacting DM model indeed predicts a stronger power suppression at small scales, as a consequence of the momentum exchange between the DM and the WZDR particles. This suppression impacts the scales probed by $\sigma_8$ (which are indicated by the light blue bands), so one would naively expect that this extension of the WZDR model can provide a simple resolution to the $S_8$ anomaly.

\begin{figure}[t]
    \centering
    \includegraphics[width=0.75\textwidth]{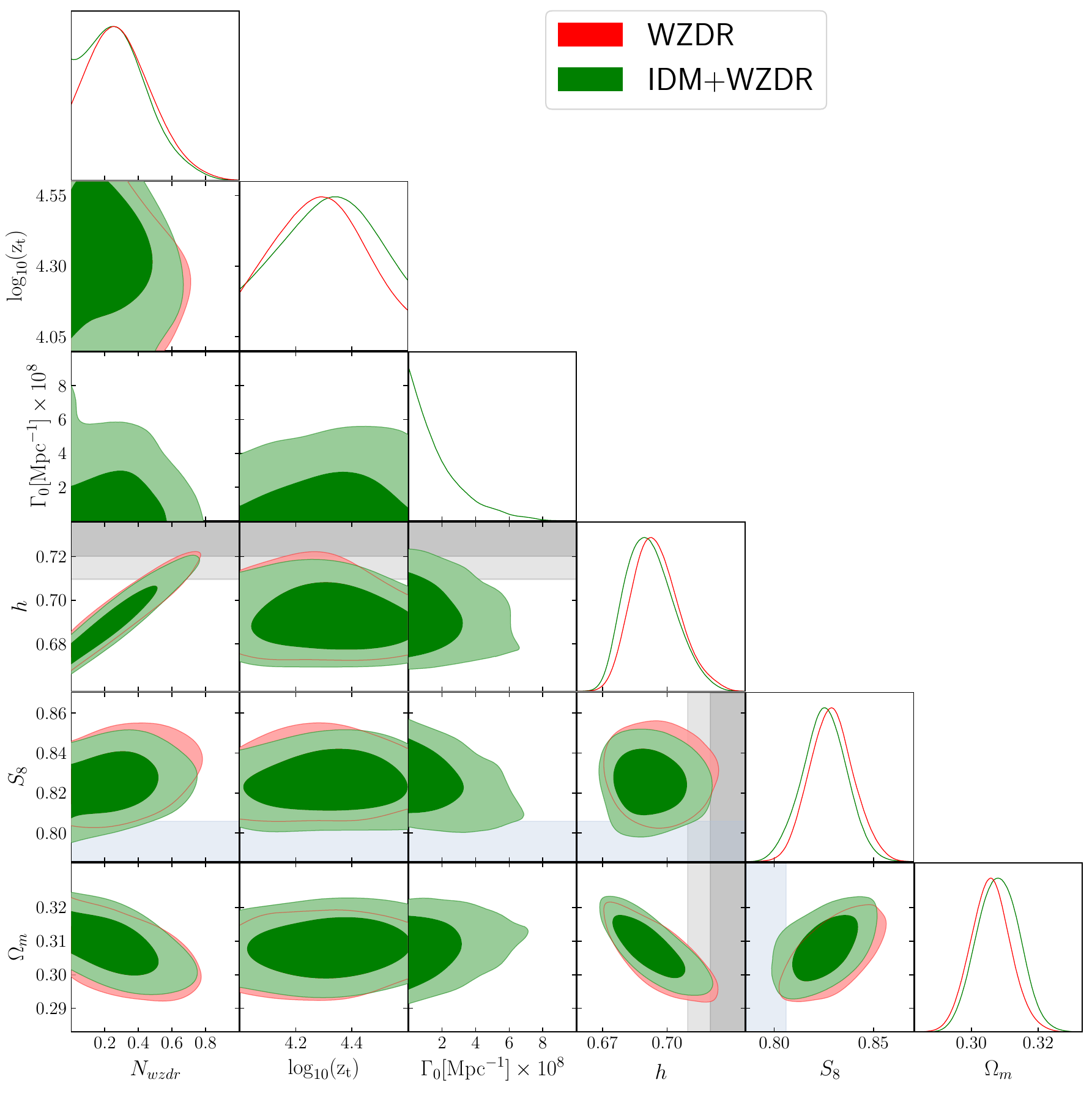}
    \caption{1D and 2D marginalized posteriors in our Planck + BAO + Pantheon analysis for the WZDR model, with and without including the presence of DM interactions. The bands are shown according to the introduction to \cref{sec:results}.} 
    \label{fig:triangle_idm_wzdr}
\end{figure}

However, we also expect a degeneracy between the abundance of WZDR $N_\mathrm{wzdr}$ and the interaction strength $\Gamma_0$, since CMB data forbids to have arbitrary amounts of interacting DR \cite{Archidiacono:2019wdp}. This is clearly seen in the top left, top right and bottom left pannels of \cref{fig:spectra_idm_wzdr}, where we report the residuals of the (lensed) TT, EE and lensing potential power spectra for the two aforementioned models. We clearly see that the IDM-WZDR model with the chosen interaction rate provides a worse fit to \emph{Planck} data than the original WZDR model. This indicates that it might not be trivial to find a region of the parameter space that solves the $S_8$ anomaly without spoiling the success in the resolution of the $H_0$ tension.

In order to check to what extent this extension of the WZDR model can provide a common resolution to both $H_0$ and $S_8$ tensions, we perform a MCMC analysis to test the IDM+WZDR model against the Planck+BAO+Pantheon data combination. We adopt unbounded flat priors on  $\{ \Omega_{\rm cdm} h^2, \Omega_b h^2, \ 100\theta_s, \ln (10^{10} A_s), \ n_s, \ \tau_{\rm reio} \}$ (note that we assume the totality of the DM to be interacting, $\Omega_{\rm cdm} = \Omega_{\mathrm{idm-wzdr}}$) and set the following priors on the three extra parameters
\begin{equation}
N_{\mathrm{wzdr}} > 0~, \ \ \ \ \ \ \mathrm{log}_{10} (z_t) \in [4, 4.6]~, \ \ \ \ \ \ \Gamma_0 [\mathrm{Mpc}^{-1}] \in [0, 10^{-7}]~.
\end{equation}

\pagebreak[20]
The results of our analysis are summarized in \cref{fig:triangle_idm_wzdr}.  We clearly see a negative correlation between $\Gamma_0$ and $N_{\rm wzdr}$, as expected. This means that the extent of the degeneracy directions in $h - N_{\rm wzdr}$ and $S_8 - \Gamma_0$ (i.e., those that are exploited in order to alleviate both tensions) is severely limited. To be more precise, the large values of $\Gamma_0$ that are required to reduce $S_8$ are not compatible with the large values of $N_{\rm wzdr}$ that are needed to increase $H_0$.

Looking at the reconstructed $S_8 = 0.82^{+0.012}_{-0.011}$ and $H_0 = 69.2^{+0.96}_{-1.3} \mathrm{km/s/Mpc}$, we see that the $S_8$ tension is only marginally alleviated, while the resolution to the $H_0$ tension is slightly worsened. We conclude that the IDM+WZDR model, despite being a natural and well-motivated extension of the WZDR model, cannot provide a common resolution to the $H_0$ and $S_8$ tensions \footnote{We have explicitely verified that letting the fraction of interacting DM $f_{\rm idm-wzdr} = \Omega_{\rm idm-wzdr}/\Omega_{\rm dm}$ free to vary (instead of fixing $f_{\rm idm-wzdr}=1$) just introduces extra degeneracies that don't help in improving the resolution to the $S_8$ tension. }.

\textbf{Comment: } Shortly after the submission of this article, two other analysis of the impact of WZDR interactions with the dark matter were released \cite{Joseph:2022jsf,Buen-Abad:2022kgf}, finding more optimistic results in terms of the $S_8$ tension. The main reason for this difference is the location of the cut-off scale in the matter power spectrum. For the scattering rate considered in our work, the DM-WZDR interactions shut off at the time of matter-radiation equality ($\log_{10}(z_{\mathrm eq}) \sim 3.5$), leading to a suppression of matter power for modes that were inside the horizon before that time, corresponding to $k \gtrsim 10^{-2} \  h/\rm{Mpc}$. This does affect the scales probed by $S_8$ but also the ones to which the CMB is very sensitive. On the contrary, the works \cite{Joseph:2022jsf,Buen-Abad:2022kgf} have derived from first principles interaction rates that shut off around the time of the step $z_t$\,, thus suppressing power only for modes that were inside the horizon at $z > z_t$, corresponding to $k \gtrsim 10^{-1} \ h/\rm{Mpc}$. 
This allows these models to reduce $S_8$ while leaving the scales mostly probed by the CMB unaffected. We notice that the shape of the matter power suppression in those two works is quite different. In \cite{Joseph:2022jsf} all of the DM is assumed to interact weakly with the DR, whereas in \cite{Buen-Abad:2022kgf} only a small fraction of the DM interacts very strongly with the DR. As a consequence, the suppression in the latter is much steeper and exhibits dark acoustic oscillations. 

However, it is worth pointing out that the main issue that we raised in \cref{sec:results_S8} is expected to hold independently of the interaction rate; namely, the fact that the CMB data imposes a negative correlation between the amount of DR and the interaction rate, making it hard to address both $H_0$ and $S_8$ tensions while simultaneously preserving a good fit to the CMB. For instance, considering the model studied in \cite{Joseph:2022jsf} and their same baseline data set, one notices a degradation in the CMB fit of $\Delta \chi^2_{\rm CMB} \sim +5$ when adding priors on $H_0$ and $S_8$ (see Tab. VIII in \cite{Joseph:2022jsf}). To check whether a similar issue could arise in the model considered in  \cite{Buen-Abad:2022kgf}, one should carry dedicated MCMC simulations. Leaving aside the problem of the $N_{wzdr}-\Gamma_0$ correlation, the differences between our work and the papers \cite{Joseph:2022jsf,Buen-Abad:2022kgf} highlight the fact that temperature dependence of the interaction rate is essential in order to determine whether this kind of models can successfully address both $H_0$ and $S_8$ tensions.

\pagebreak[20]
\section{Conclusions} \label{sec:concl}

We have investigated the WZDR model proposed in \cite{Aloni:2021eaq} as a candidate for easing the Hubble tension. We have developed a slightly more general framework for tracking the cosmological evolution and applied it in order to investigate the fundamental assumptions made in the original analysis of \cite{Aloni:2021eaq}. We have shown that the precise assumption on particle properties (spin-statistics, abundances) has only a very small impact on the overall conclusions. Instead, the width of the prior for a Bayesian analysis does appear to have a mild impact, as expected due to prior volume effects. The impact of BBN on this model is ignored in \cite{Aloni:2021eaq}, where the authors argue that the dark species could have been produced only after BBN, requiring a further extension of the model. If one takes into account the impact of the BBN constraints on such a model, its ability to ease the Hubble tension is severely limited. 

We also find that the model slightly exasperates the $A_L$ tension (2.1$\sigma$) compared to $\Lambda$CDM (1.9$\sigma$), while the preference for negative $\Omega_k$ is almost unchanged. Furthermore, the WZDR does significantly ease the tension between parameter inference from the temperature autocorrelation for $\ell < 800$ and $\ell \geq 800$ (from $\sim 1.8\sigma$ to $0.5\sigma$ for the $\Omega_m$ parameter only).

When confronted with combinations of different/additional data (BAO+BBN, WMAP, ACT, SPT) no strong preference for the model is found in any of the investigated cases (unlike for EDE \cite{Poulin:2021bjr,Smith:2022hwi}, where for example the combination of WMAP+ACT prefers a presence of EDE). However, the tension remains mostly eased. Only in the case of combining ACT or SPT with Planck 2018 TT+TE+EE ($\ell < 1050$) data, the abilities of the model to ease the tension is more severely constrained. We show that this is mostly due to the excellent constraining power of the polarization data on models of additional dark radiation. Models that rely on the inclusion of dark radiation to ease the Hubble tension will necessarily have to address the increasing constraining power of the CMB polarization measurements in order to be successful.

\enlargethispage*{2\baselineskip}
We find no significant impact on the $S_8$ tension either from any of these combinations when assuming a temperature-independent scattering. Even when extended with dark-radiation dark-matter interactions, we find that the abundance of the dark radiation is too constrained for its interaction to significantly impact the $S_8$ tension. On the other hand, other analyses such as \cite{Joseph:2022jsf,Buen-Abad:2022kgf} have shown that a well-motivated temperature-dependent interaction rate can help easing the $S_8$ tension. Thus, while the WZDR model presents a thoroughly interesting entry in the list of viable proposals to ease the Hubble tension, we neither find a preference for this model in any existing data (not including a $H_0$ prior from the local distance ladder) nor any additional capabilities to ease other tensions even when allowing for additional temperature-independent dark-radiation dark-matter interactions.

\section*{Acknowledgements}
The authors would like to thank Martin Schmaltz, Melissa Joseph, and Daniel Aloni for their frequent interactions, useful advice, and cross-checks. The authors would also like to thank Julien Lesgourgues and Sven Günther for their continued support and advice, as well as their contributions in the writing stage of this draft. The authors would also like to thank Licia Verde and Vivan Poulin for their useful comments.
Nils Sch\"oneberg acknowledges the support of the following Maria de Maetzu fellowship grant: Esto publicaci\'on es parte de la ayuda CEX2019-000918-M, financiado por MCIN/AEI/10.13039/501100011033.
\appendix
\section{Thermodynamic derivations}\label{app:proof_R}

We remind ourselves of the first law of thermodynamics
\begin{equation}
    \mathrm{d}U = T \mathrm{d}\tilde{S} - P \mathrm{d}V + \sum_i \mu_i \mathrm{d}N_i~,
\end{equation}
which holds separately for the dark sector (which is in local thermodynamic equilibrium), since it holds also for the visible sector. By noticing $U = \rho V$ and $\tilde{S} = S V$ and $V \propto a^3$ (at zeroth order in perturbations), as well as $\mu_\xi = \mu_\phi = 0$ for the new dark species, we find
\begin{equation}
    \mathrm{d} (\rho a^3) = T \mathrm{d} (Sa^3) - P \mathrm{d}(a^3)~.\label{app:eq:firstlaw}
\end{equation}
Together with the energy density conservation equation\footnote{This is derived directly from the twice reduced Bianchi identity in the homogeneous FLRW metric, and does only refer to the total pressure and density. Since, however, the pressures and densities of the species of the standard cosmological model separately obey these equations as well, they are independently valid also for the \textit{total} of the dark sector. A more common form is $\mathrm{d}\rho/\mathrm{d}t = -3 H(\rho+P)$, which is directly equivalent to $\mathrm{d}\rho = -3 (\rho+P)\mathrm{d}\ln a$ due to $H = \mathrm{d}\ln a/\mathrm{d}t$ and together with $\mathrm{d}(a^3) = 3a^3 \mathrm{d}\ln a$ this directly gives $a^3\mathrm{d}\rho = -(\rho+P) \mathrm{d}(a^3)$, which can be easily put into the given form of \cref{app:eq:energycons}}
\begin{equation}
    \mathrm{d} (\rho a^3) = - P \mathrm{d}(a^3)~,\label{app:eq:energycons}
\end{equation}
this implies $\mathrm{d}(Sa^3) = 0$ independently of the temperature $T$. As such, another equation is required to determine the evolution of $T(t)$. As we will show below, this is the density conservation \cref{app:eq:energycons}.

We now use $\mathrm{d} \rho = - 3(\rho+P) \mathrm{d}\ln a$ (which can be derived directly from \cref{app:eq:energycons}) to determine that
\begin{equation}
    \der{\ln T}{\ln a} = \frac{-3(\rho+P)}{\mathrm{d}\rho/\mathrm{d}\ln T}~.\label{app:eq:temperature_partial}
\end{equation}
The temperature derivative of the energy density of both dark species can be computed explicitly. For the massless species the result is simply $\mathrm{d}\rho_\phi/\mathrm{d}\ln T_\phi = 4\rho_\phi$\,, while for the massive species one needs to explicitly perform the partial integration
\begin{align}
    \der{\rho_\xi}{\ln T} & = \frac{g_\xi}{2\pi^2}\int p^2 E \left(\frac{\partial f_\xi(E,t)}{\partial \ln T} \right)\Bigg|_{E} \mathrm{d}p = -\frac{g_\xi}{2\pi^2}\int p^2 E^2 \left(\frac{\partial f_\xi(E,t)}{\partial E} \right) \Bigg|_{\ln T} \mathrm{d}p \\ & = -\frac{g_\xi}{2\pi^2}\int p E^3 \left(\frac{\partial f_\xi(E,t)}{\partial E} \right) \Bigg|_{\ln T} \mathrm{d}E = \frac{g_\xi}{2\pi^2}\int \frac{\partial [p E^3]}{\partial E} f_\xi(E,t) \mathrm{d}E \\ &= \frac{g_\xi}{2\pi^2}\int (3 E^2 p + E^4/p) f_\xi(E,t) \mathrm{d}E = \frac{g_\xi}{2\pi^2}\int (3 E p^2 + E^3) f_\xi(E,t) \mathrm{d}p \\ &= 3 \rho_\xi + 3 R_\xi~, \label{eq:rhoevolv}
\end{align}
with the pseudo-density $R_\xi$ as introduced in \cref{eq:R}. In the first line we have used $f(E,t)=f(E/T) = f(\exp[\ln E - \ln T])$ to write $\partial f/\partial \ln T = - \partial f/\partial \ln E$. In fact, since the derivation assumes nothing more than this $f(E,t) = f(E/T)$, it holds true even for the massless species $\phi$ for which $\rho_\phi = 3R_\phi$ (trivially from setting $E=p$ in \cref{eq:rho,eq:R}) and thus $\mathrm{d}\rho_\phi/\mathrm{d}\ln T = 4 \rho_\phi$ as expected since $\rho_\phi \propto T_\phi^4$ \,. 
We can thus write even for the total energy density the remarkably simple
\begin{equation}
    \der{\rho}{\ln T} = 3 (\rho + R)~.
\end{equation}
with the understanding that $R = R_\xi + R_\phi = R_\xi + \rho_\phi/3$\,. Together with \cref{app:eq:temperature_partial} one can directly show that
\begin{equation}
    \der{\ln T}{\ln a} = - \frac{\rho+P}{\rho+R}~,
\end{equation}
as used in \cref{eq:Tevolv}. Finally, for the purpose of determining the sound speed of the dark species we additionally want to determine $\mathrm{d}P/\mathrm{d}\ln a$\,. This can be done analogously to the derivation for $\der{\rho}{\ln T}$ and leads to 
\begin{equation}
    \der{P}{\ln T} = (\rho+P)~.\label{app:eq:Pevolv}
\end{equation}
Indeed, this fact could also be shown using the Gibbs-Duhem relation (the requirement that the total energy in the first law of thermodynamics is a proper integrable differential, which directly comes from Stokes law) which states that
\begin{equation}
    0 = a^3 S \mathrm{d}T - a^3 \mathrm{d}P~. \label{app:eq:gibbs}
\end{equation}
This equation added onto \cref{app:eq:firstlaw} also allows one to write the well known equation for the entropy $T S = (\rho +P)$ applicable for negligible chemical potential. Together with \cref{app:eq:gibbs} this allows us to determine \cref{app:eq:Pevolv} directly.

Using either derivation, the evolution of the pressure by itself does not provide further information. Instead, it can be used to determine the (adiabatic) sound speed since 
\begin{equation}
    c_s^2 = \der{P}{\rho} = \frac{\mathrm{d}P/\mathrm{d}\ln T}{\mathrm{d}\rho/\mathrm{d}\ln T} = \frac{\rho+P}{3(\rho+R)}~. \label{app:eq:cs2}
\end{equation}
We notice that for relativistic species $R = P$ and thus $c_s^2 = 1/3$ as expected.

As a small side-note we want to point out that the derivation for $\mathrm{d}\rho/\mathrm{d}\ln T$ is very similar to that for massive neutrinos, which leads to the definition of a pseudo-pressure. However, unlike in that case, we can not assume $T \propto a^{-1}$ as for a decoupled species, which naturally leads to different equations of motion.

\section{Equivalence to previous work}

To establish the equivalence of the derived formalism under the assumption of the Maxwell Boltzmann distribution, we can explicitly perform the integrals of \cref{eq:rho,eq:P,eq:R} to find
\begin{align}
    \rho_\xi &= \rho_B \left[\frac{x^2}{2}K_2(x) + \frac{x^3}{6}K_1(x)\right]~,\\
    P_\xi &= \frac{\rho_B}{3} \left[\frac{x^2}{2}K_2(x) \right]~,\\
    R_\xi &= \left(1+\frac{x^2}{3}\right) \cdot P_\xi~,
\end{align}
with $\rho_B = \frac{\pi^2}{30} g_\xi T_\xi^4$ and $x = m_\xi/T_\xi$\,. Reminding oneself that $\rho_B \propto x^{-4}$, one can quickly find that
\begin{align}
    \der{P_\xi}{\ln T} &= - \der{[x^{-4} \cdot x^2/2 \cdot K_2(x)]}{\ln x} \cdot (\rho_B x^4) = \frac{x^3}{6} K_3(x) \rho_B = \rho_\xi + P_\xi~, \\
    \der{\rho_\xi}{\ln T} &= - \der{[x^{-4} \cdot (\frac{x^2}{2}K_2(x) + \frac{x^3}{6}K_1(x))]}{\ln x} \cdot (\rho_B x^4) \\ & = \frac{x}{6} \left[3 x^2 K_1(x) + (12x+x^3) K_2(x)\right] \rho_B \\
    & = 3 (\rho_\xi + R_\xi)~,
\end{align}
thus confirming that the conclusion of \cref{eq:rhoevolv,app:eq:Pevolv} holds also in the Maxwell-Boltzmann distributed case. We can further check that equation (A10) of \cite{Aloni:2021eaq} also holds in this generalized formalism. First, we recognize it is equivalent to
\begin{equation}
    (x a_t/a)^3 = 1 + \frac{3}{4} (\rho_\xi + P_\xi)/\rho_\phi =: f(x)~,
\end{equation}
with $a_t$ defined in \cref{eq:zt}. Now we can notice that $\frac{4}{3} f(x) \rho_\phi =(\rho_\phi+P_\phi + \rho_\xi + P_\xi) = \rho + P$. Then, we multiply the above equation by $\frac{4}{3} \rho_\phi$, take the logarithm, and finally derive with respect to $\ln a$ to give
\begin{equation}
    -3 + 3 \der{\ln x}{\ln  a} + \der{\ln \rho_\phi}{\ln x} \der{\ln x}{\ln a}= \der{\ln (\rho+P)}{\ln x} \der{\ln x}{\ln a}~.
\end{equation}
Now since $\der{\rho_\phi}{\ln x} = -4 \rho_\phi$ one can very easily solve for the derivative of $\ln x$ to find
\begin{equation}
    \der{\ln T}{\ln a} = - \der{\ln x}{\ln a} = \frac{-3}{1+\der{\ln (\rho+P)}{\ln x}} = -\frac{(\rho+P)}{\frac{1}{3}(\rho+P)-\frac{1}{3}\der{(\rho+P)}{\ln T}}~.
\end{equation}
With the aformentioned facts and $R_\phi = P_\phi = \rho_\phi/3 \propto T^4$ it is easy to conclude that 
\begin{equation}
    \der{(\rho+P)}{\ln T} = 3 (\rho+R) + (\rho+P)~,
\end{equation}
which indeed does give the correct evolution equation of
\begin{equation}
    \der{\ln T}{\ln a} = - \frac{\rho+P}{\rho+R}~,
\end{equation}
as already mentioned in \cref{eq:Tevolv}. We conclude that this generalized approach must, in principle, exactly recover the evolution of \cite{Aloni:2021eaq} in the limit of $\varepsilon_\xi = 0$ of \cref{eq:distribution} (Maxwell-Boltzmann distribution).

\bibliographystyle{JHEP}
\bibliography{WZDR.bib}
\end{document}